\documentclass[landscape,onecolumn,letterpaper]{article}
\usepackage{epsfig}
\newpage

\begin{document}
\begin{figure}
\begin{center}
\epsfig{file=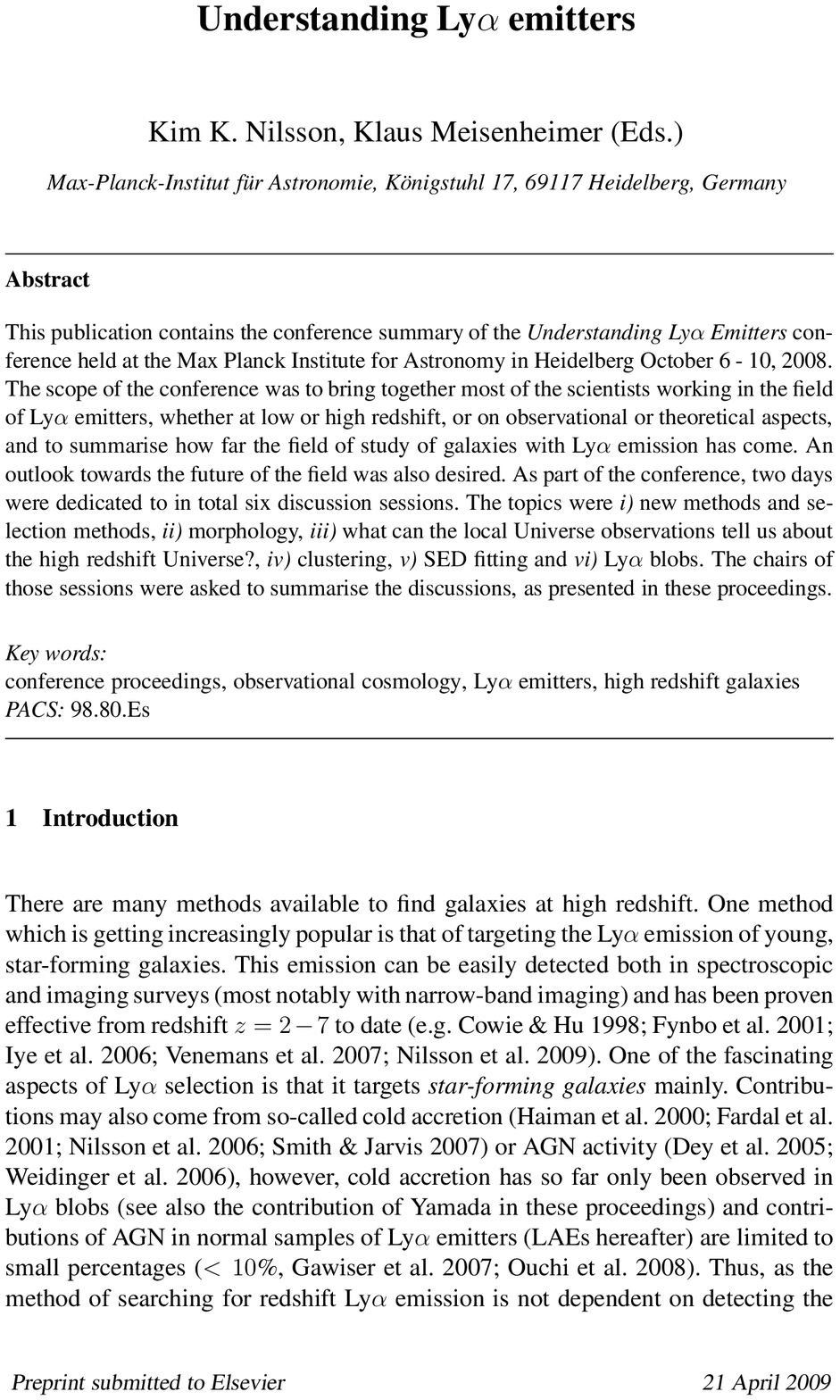,width=25cm,clip=}
\end{center}
\end{figure}
\newpage
\begin{figure}
\begin{center}
\epsfig{file=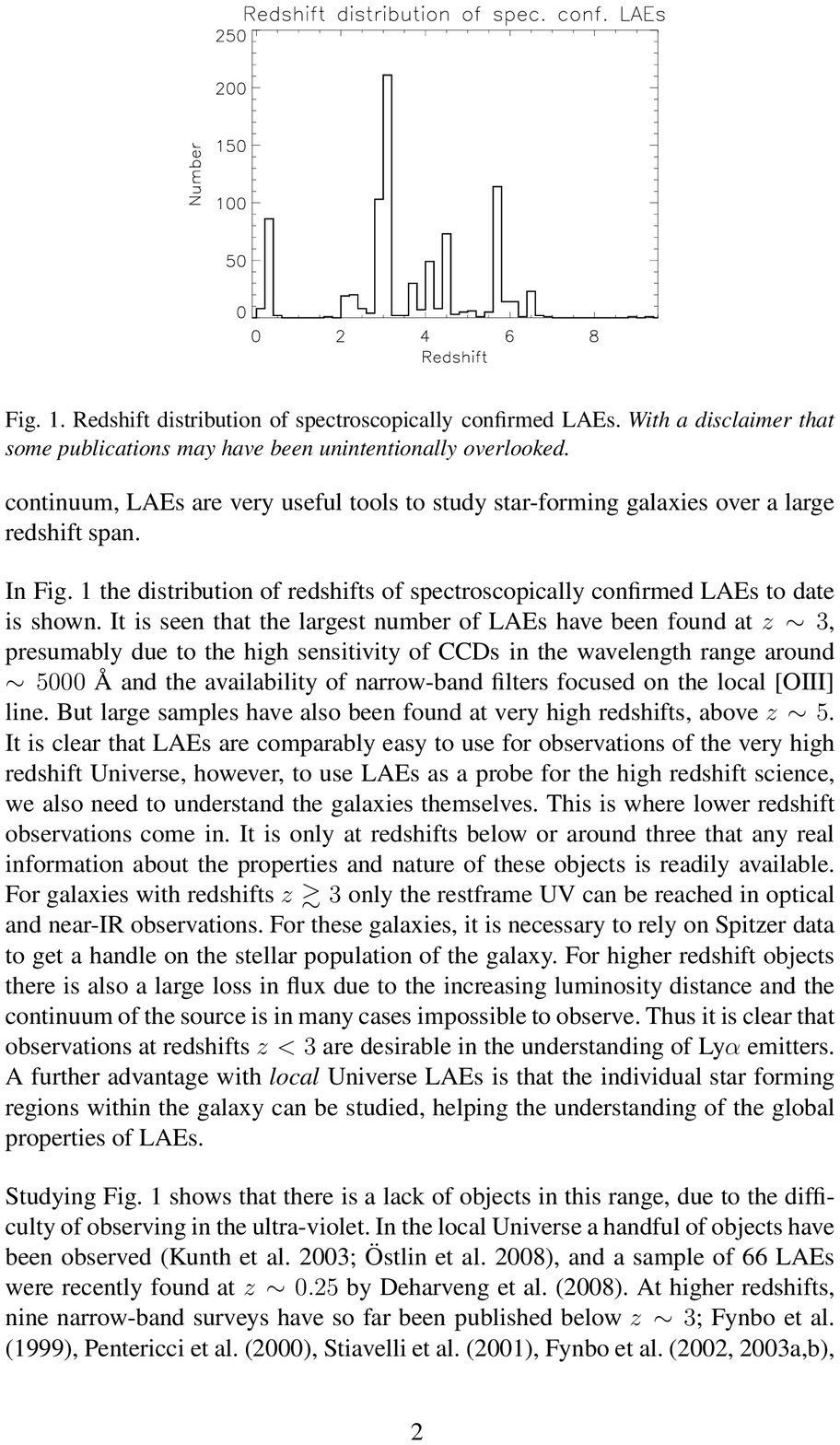,width=25cm,clip=}
\end{center}
\end{figure}
\newpage
\begin{figure}
\begin{center}
\epsfig{file=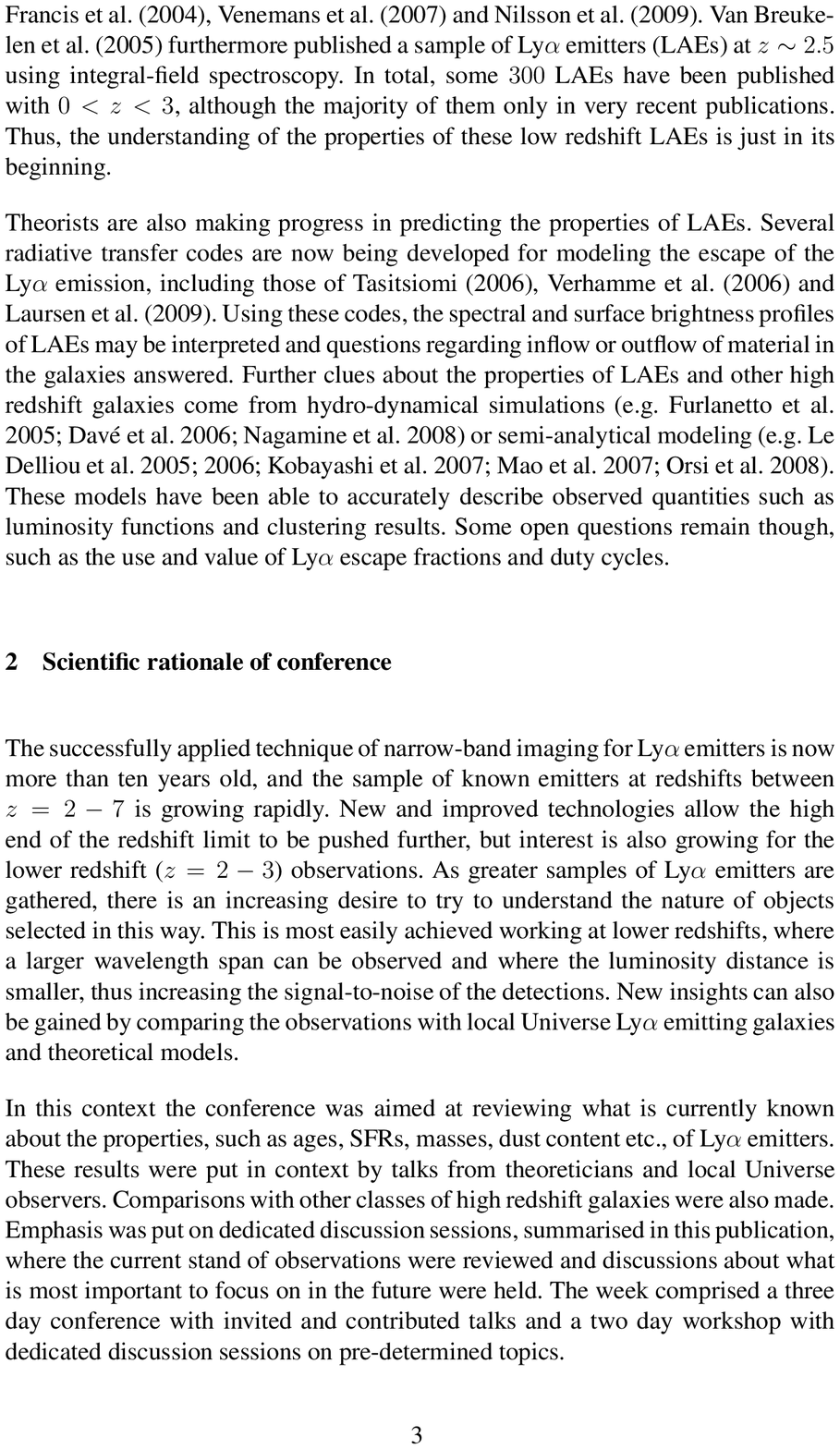,width=25cm,clip=}
\end{center}
\end{figure}
\newpage
\begin{figure}
\begin{center}
\epsfig{file=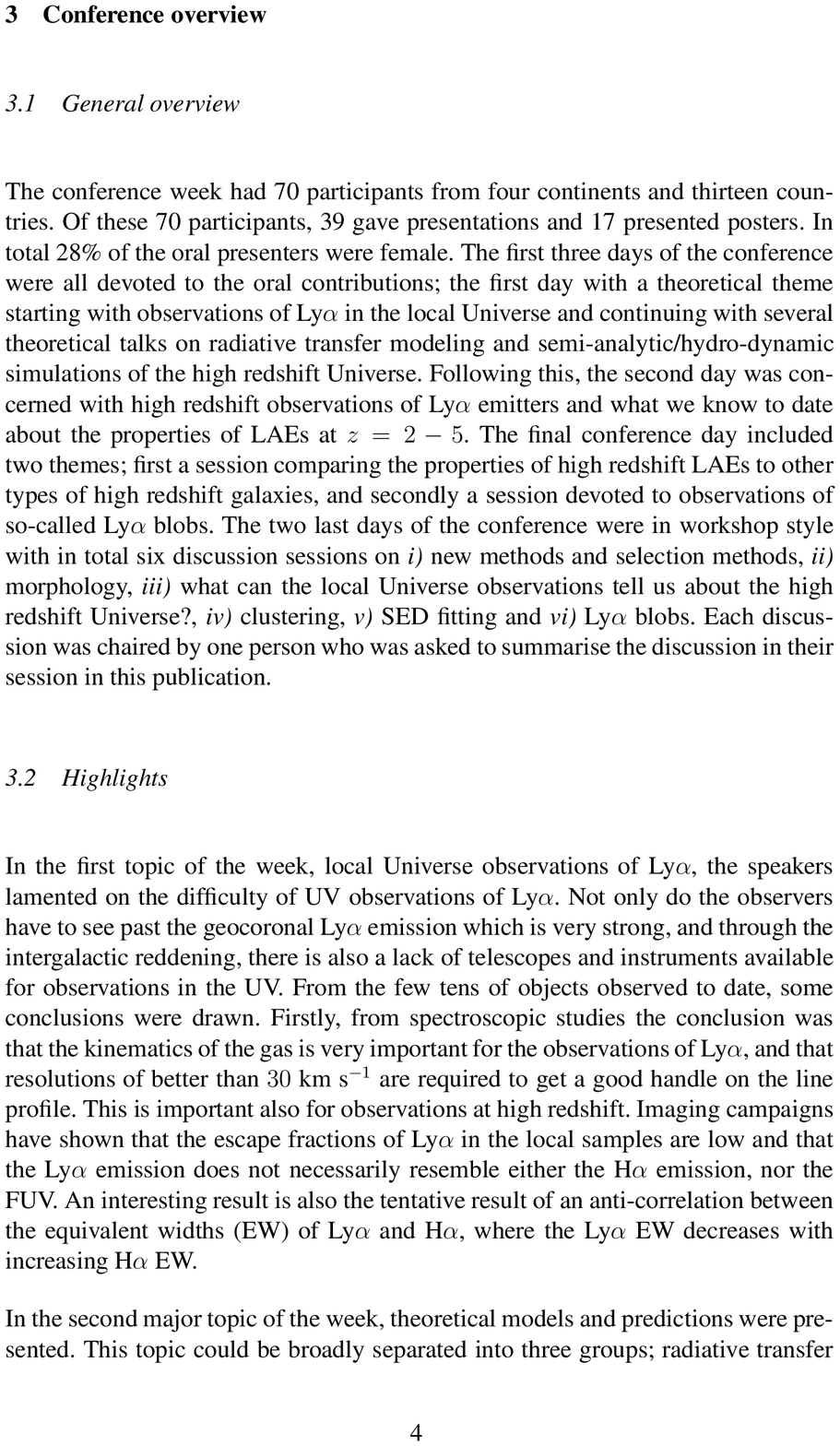,width=25cm,clip=}
\end{center}
\end{figure}
\newpage
\begin{figure}
\begin{center}
\epsfig{file=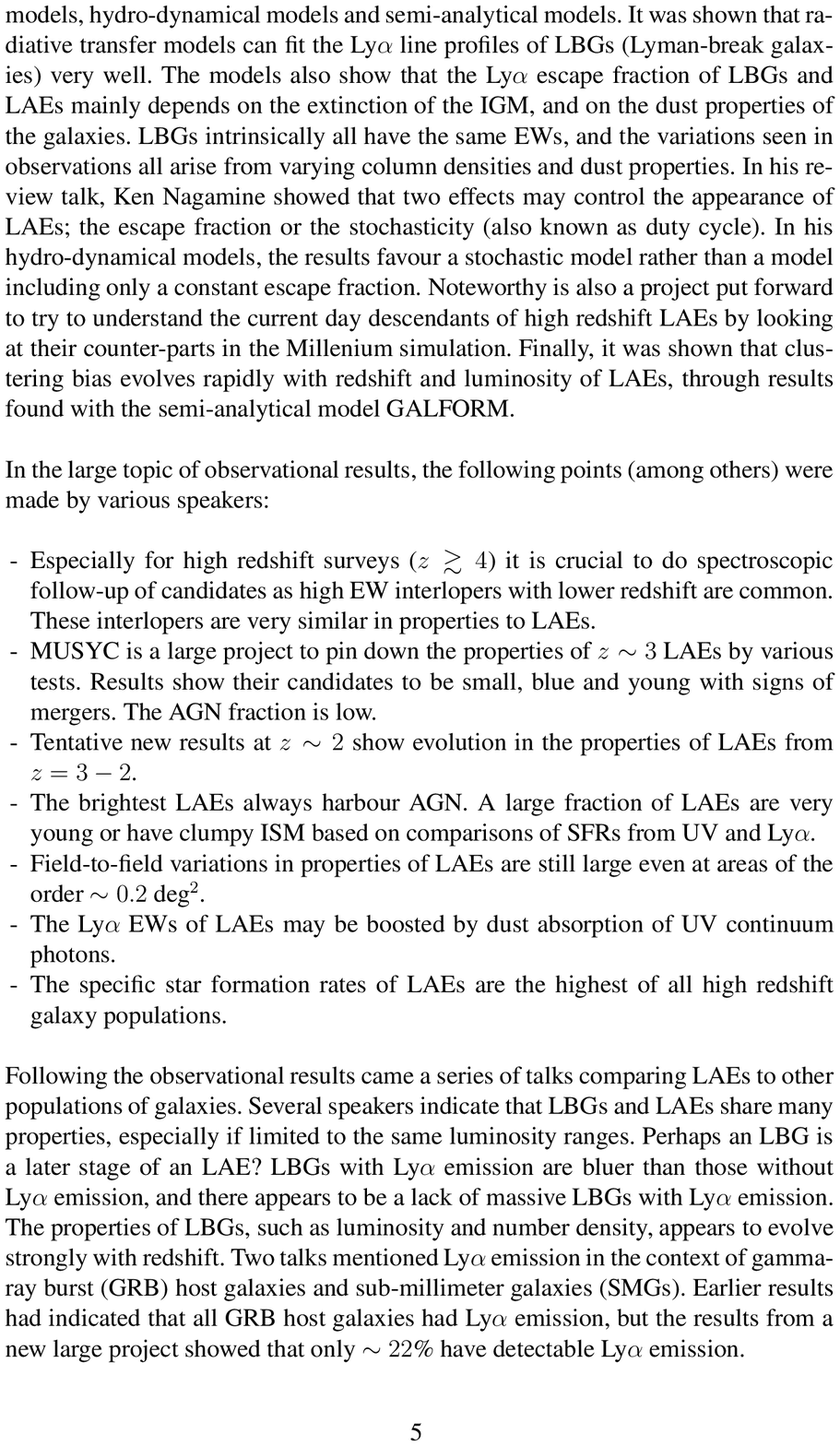,width=25cm,clip=}
\end{center}
\end{figure}
\newpage
\begin{figure}
\begin{center}
\epsfig{file=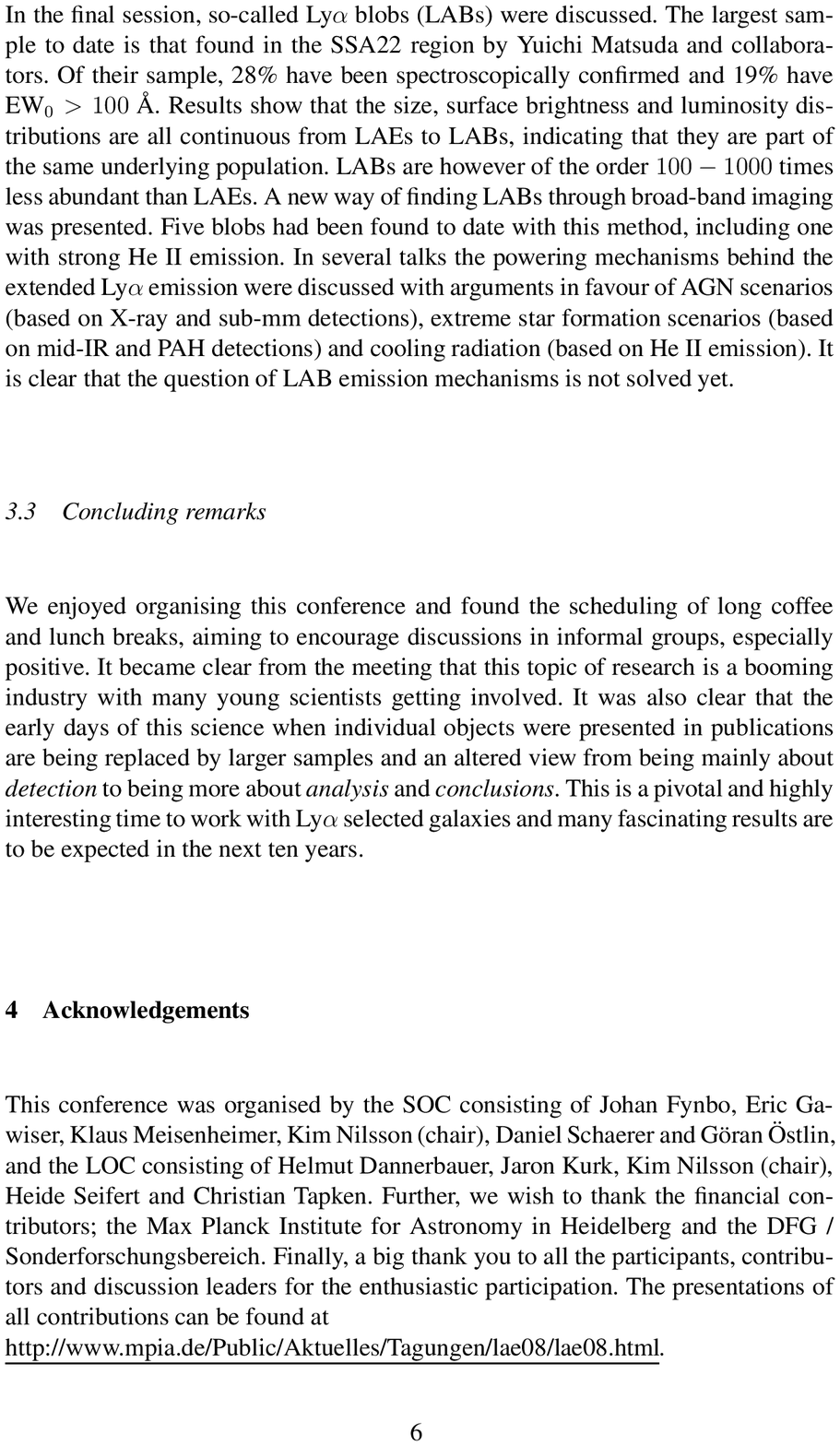,width=25cm,clip=}
\end{center}
\end{figure}
\newpage
\begin{figure}
\begin{center}
\epsfig{file=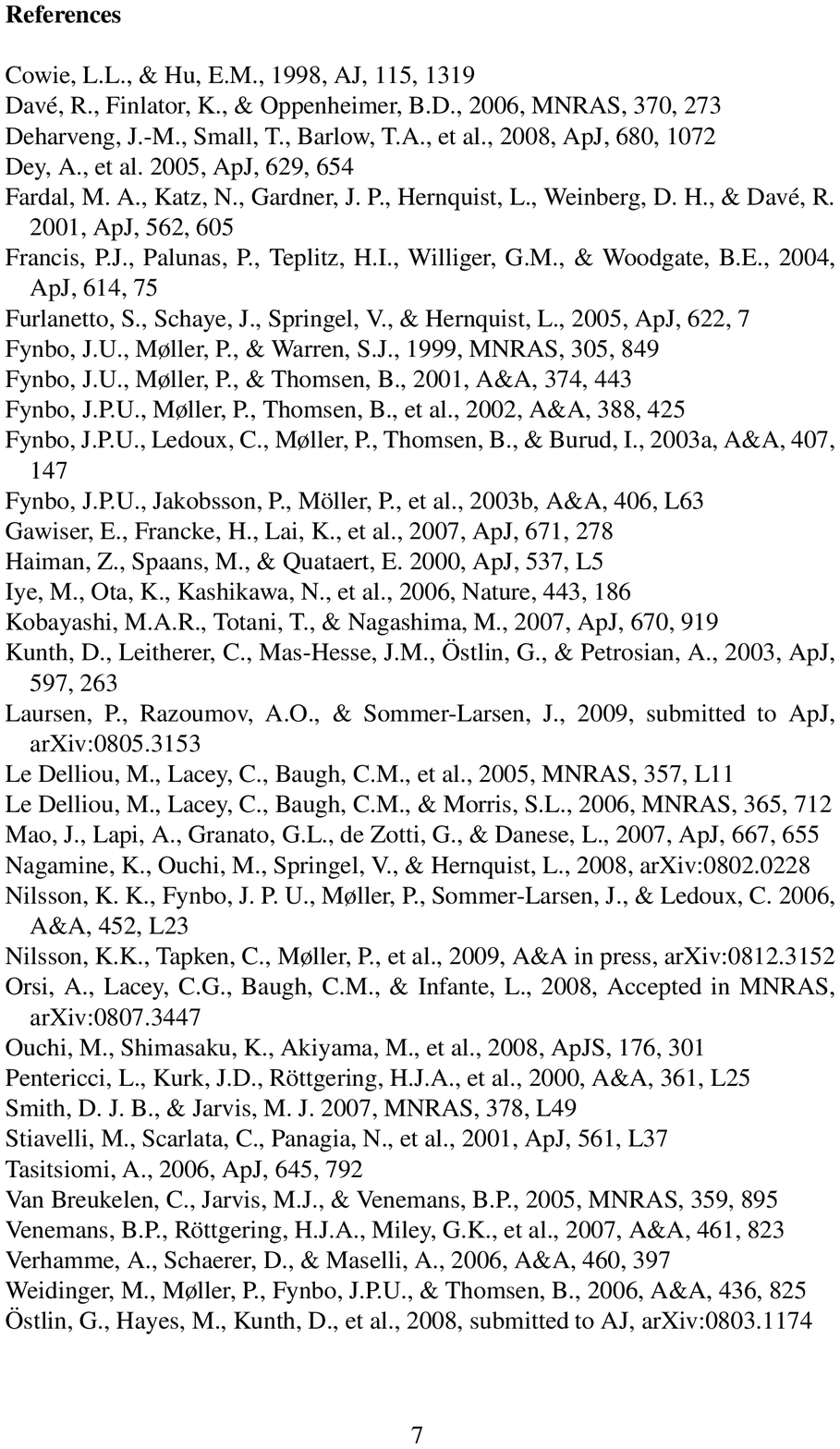,width=25cm,clip=}
\end{center}
\end{figure}
\begin{figure}
\begin{center}
\epsfig{file=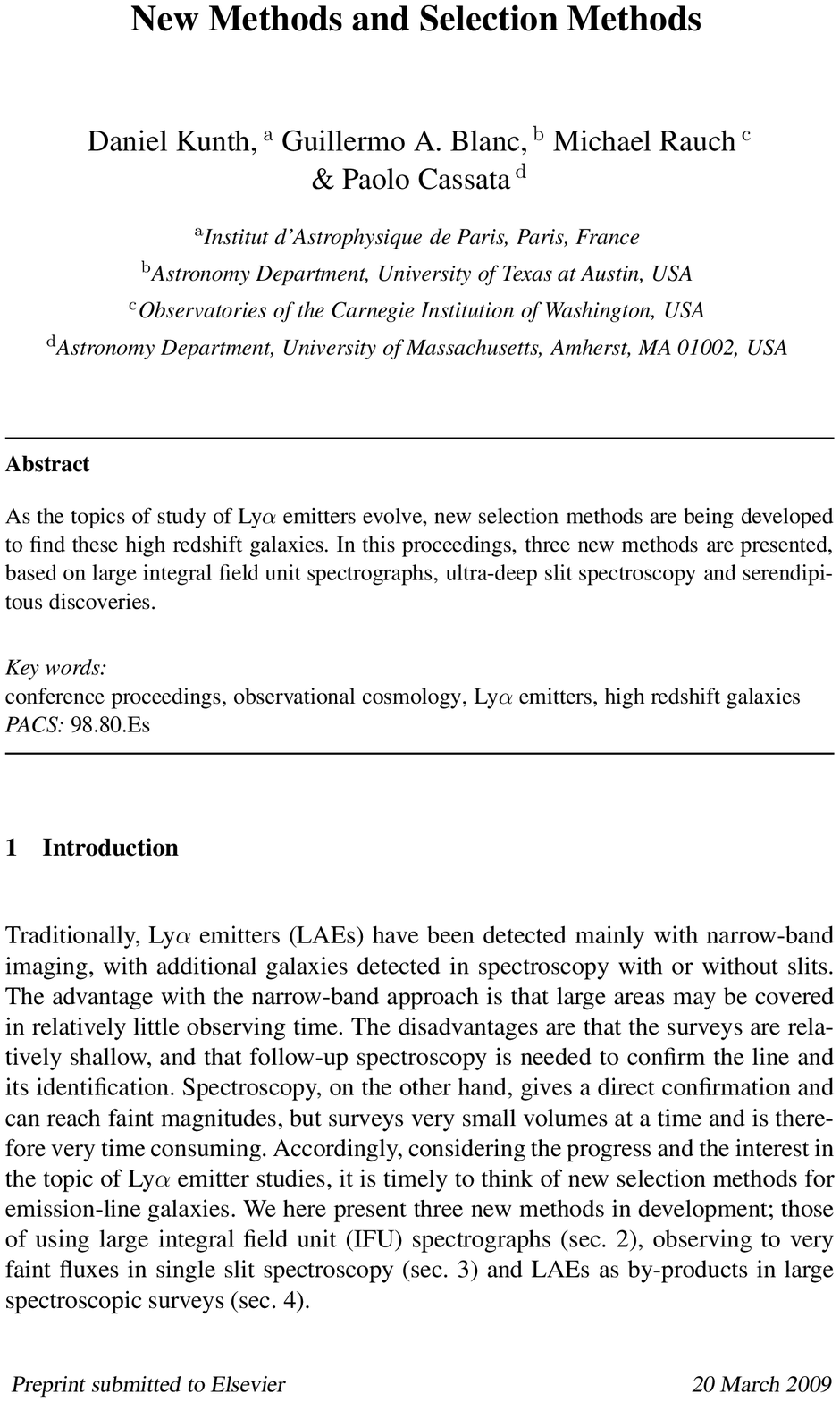,width=25cm,clip=}
\end{center}
\end{figure}
\newpage
\begin{figure}
\begin{center}
\epsfig{file=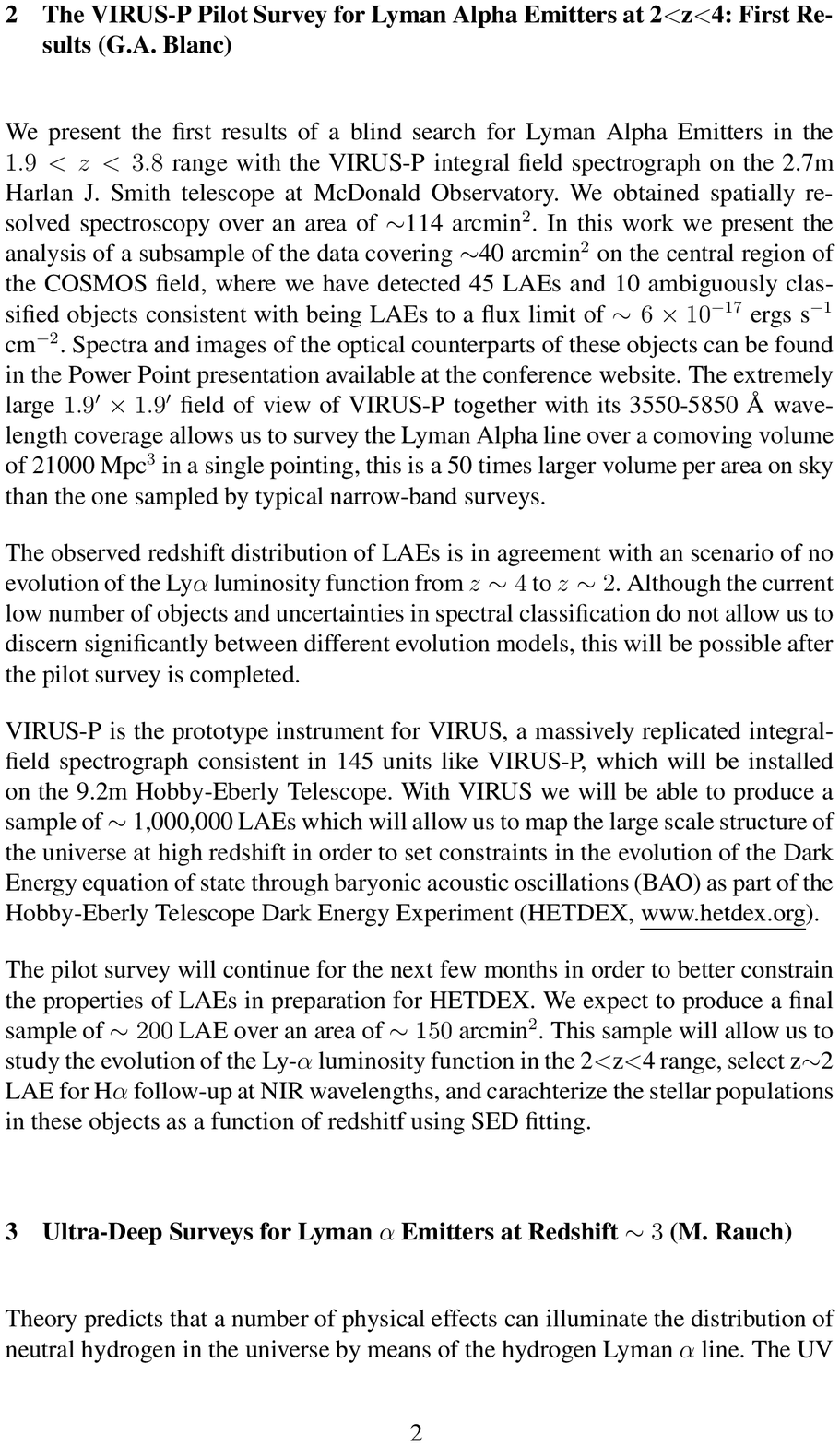,width=25cm,clip=}
\end{center}
\end{figure}
\newpage
\begin{figure}
\begin{center}
\epsfig{file=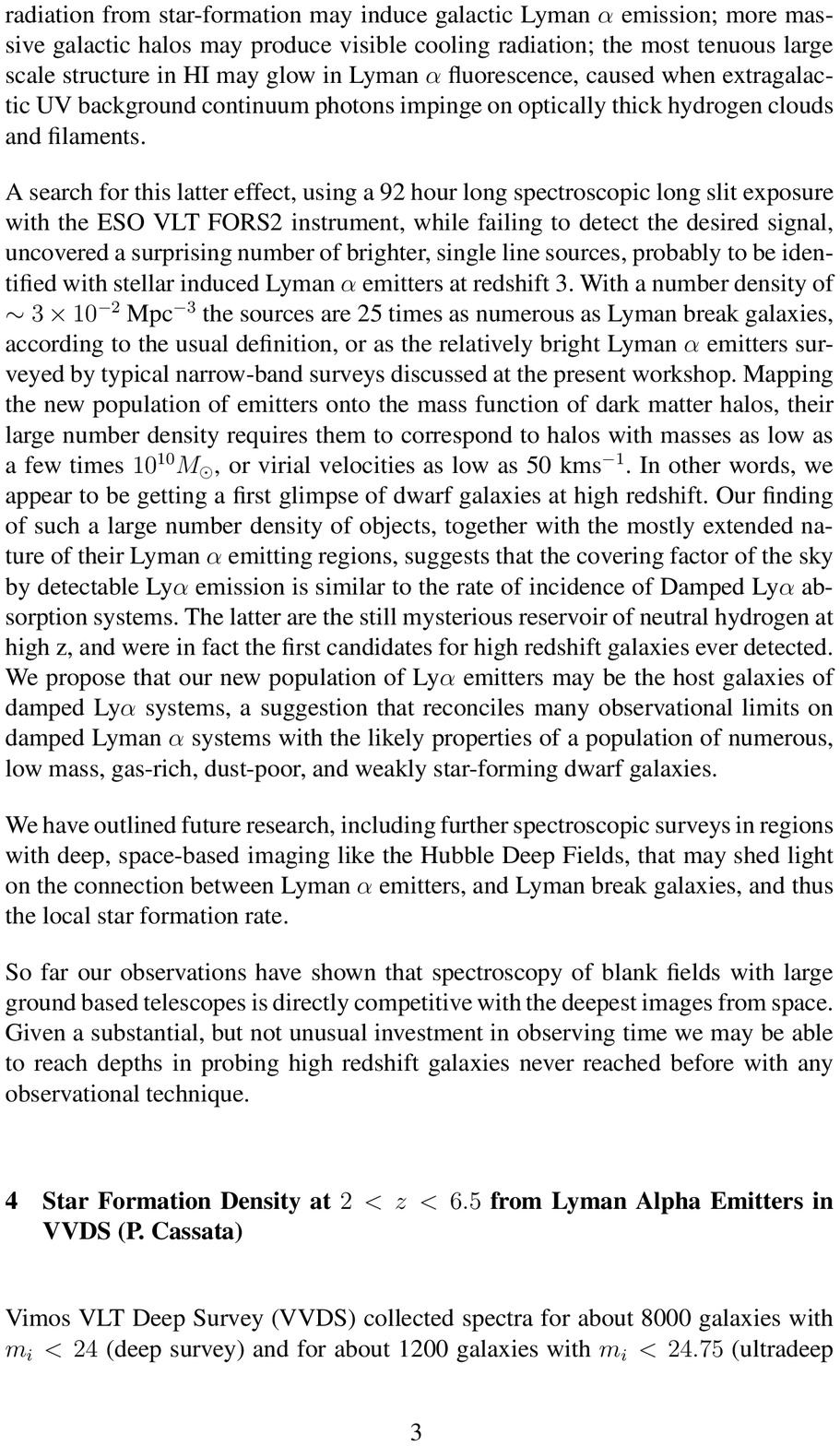,width=25cm,clip=}
\end{center}
\end{figure}
\newpage
\begin{figure}
\begin{center}
\epsfig{file=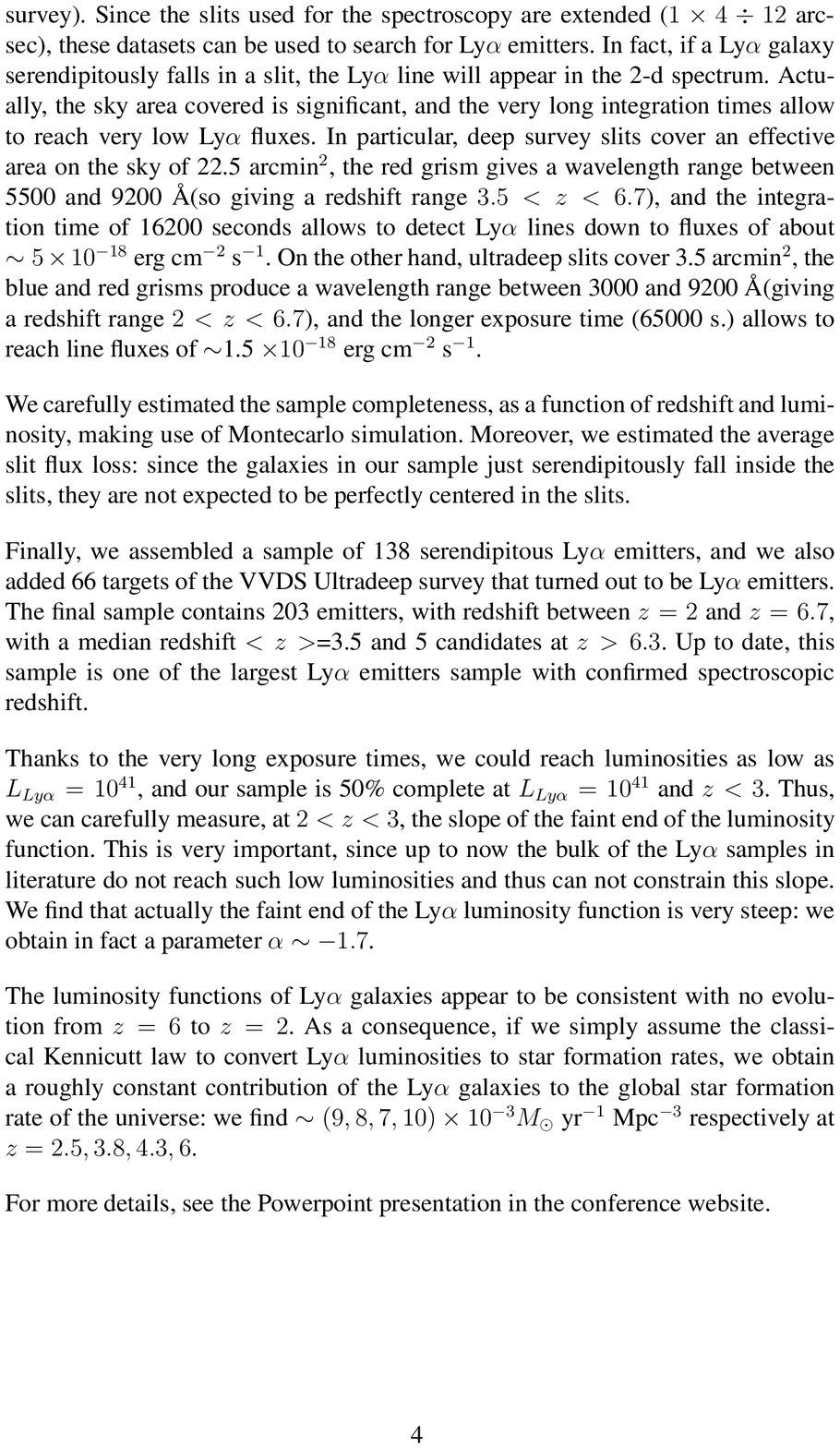,width=25cm,clip=}
\end{center}
\end{figure}
\begin{figure}
\begin{center}
\epsfig{file=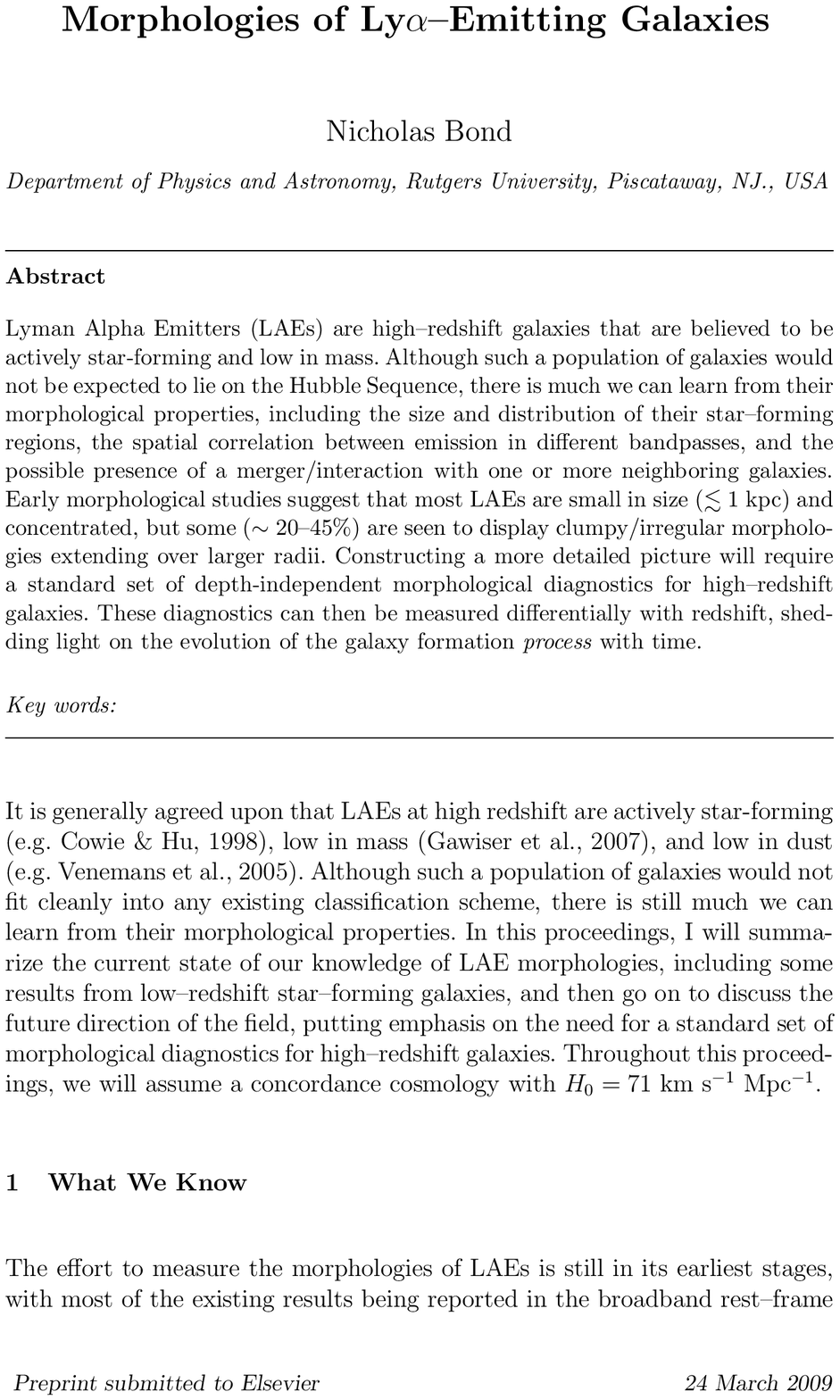,width=25cm,clip=}
\end{center}
\end{figure}
\newpage
\begin{figure}
\begin{center}
\epsfig{file=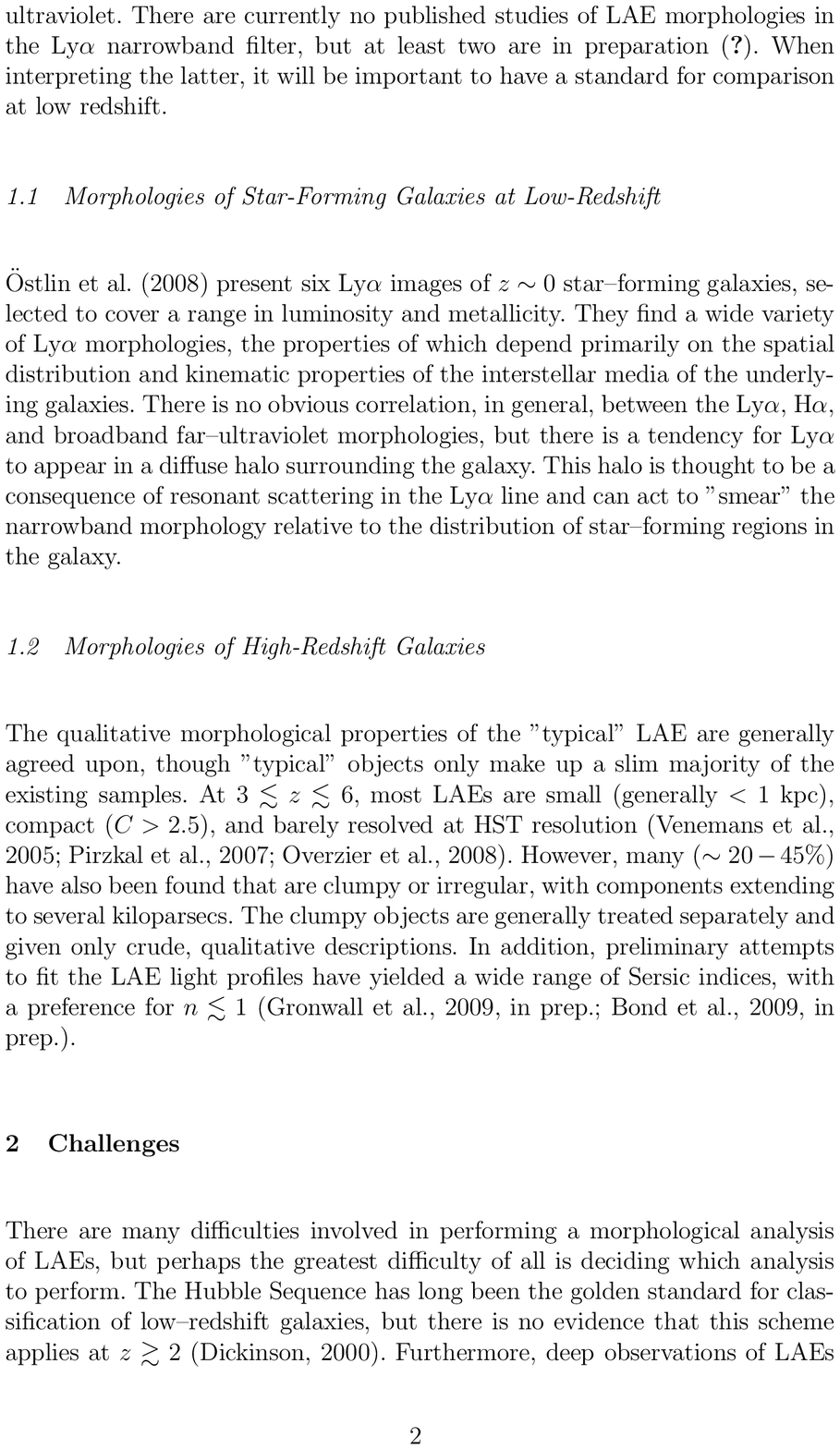,width=25cm,clip=}
\end{center}
\end{figure}
\newpage
\begin{figure}
\begin{center}
\epsfig{file=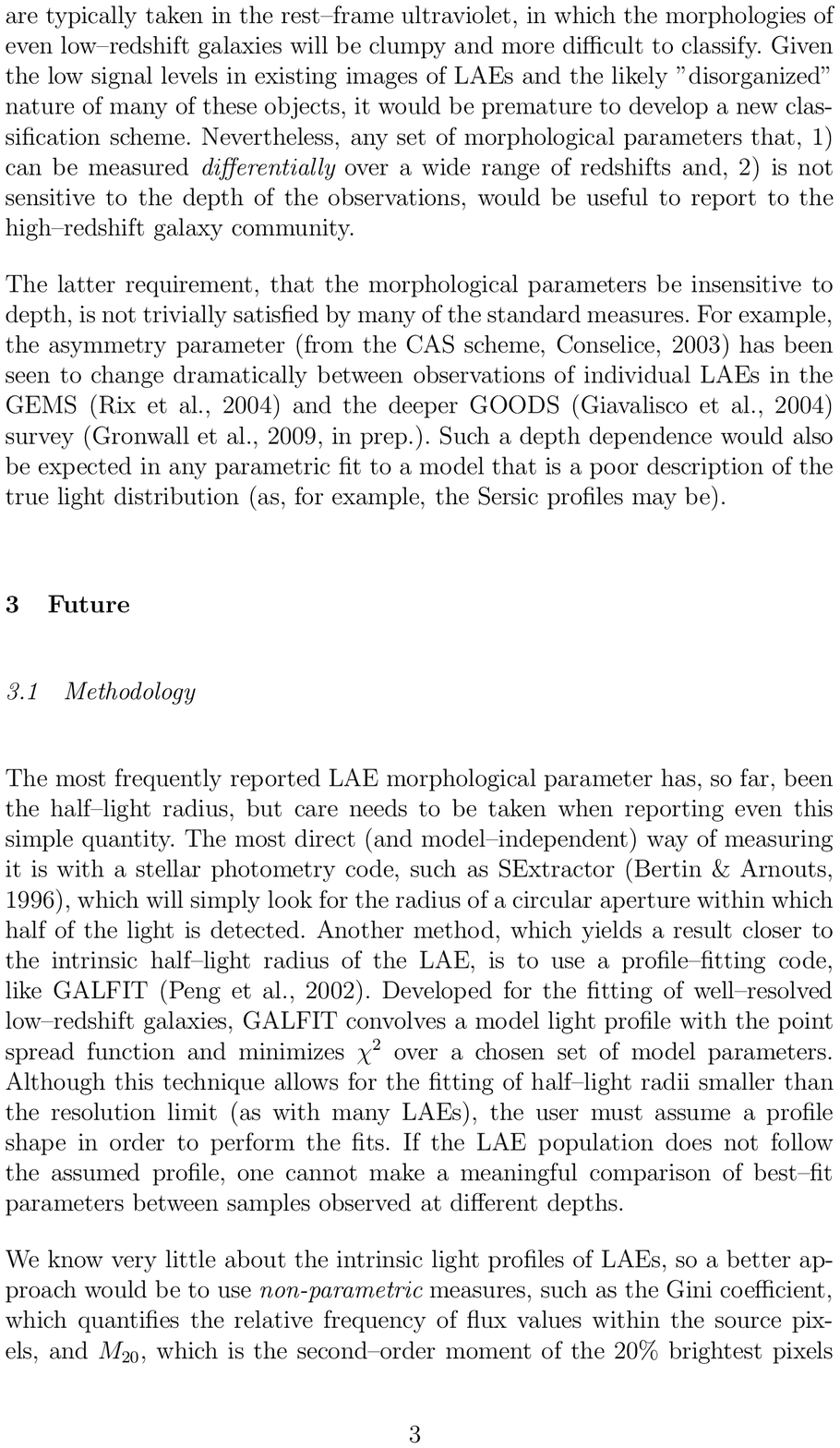,width=25cm,clip=}
\end{center}
\end{figure}
\newpage
\begin{figure}
\begin{center}
\epsfig{file=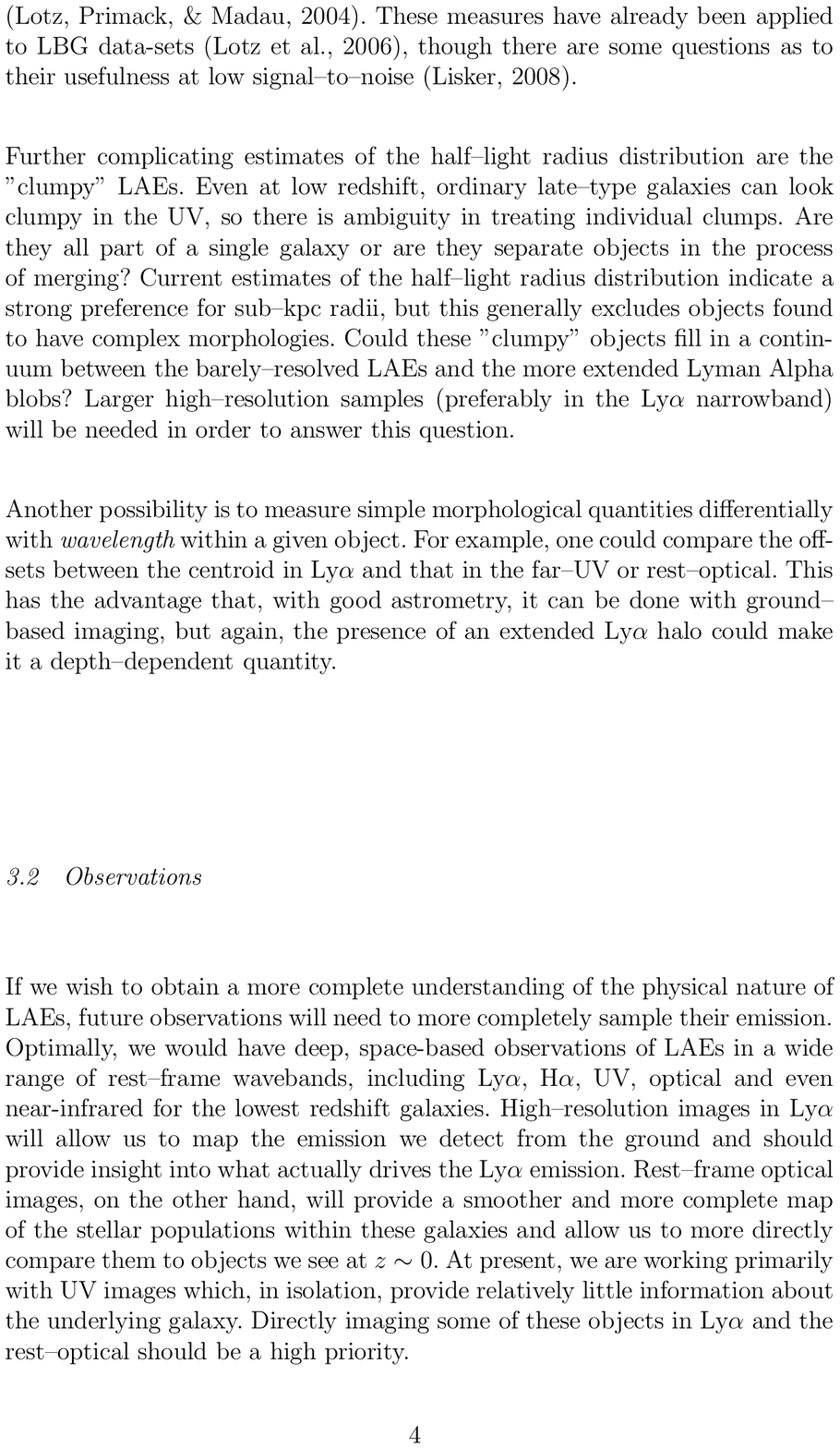,width=25cm,clip=}
\end{center}
\end{figure}
\newpage
\begin{figure}
\begin{center}
\epsfig{file=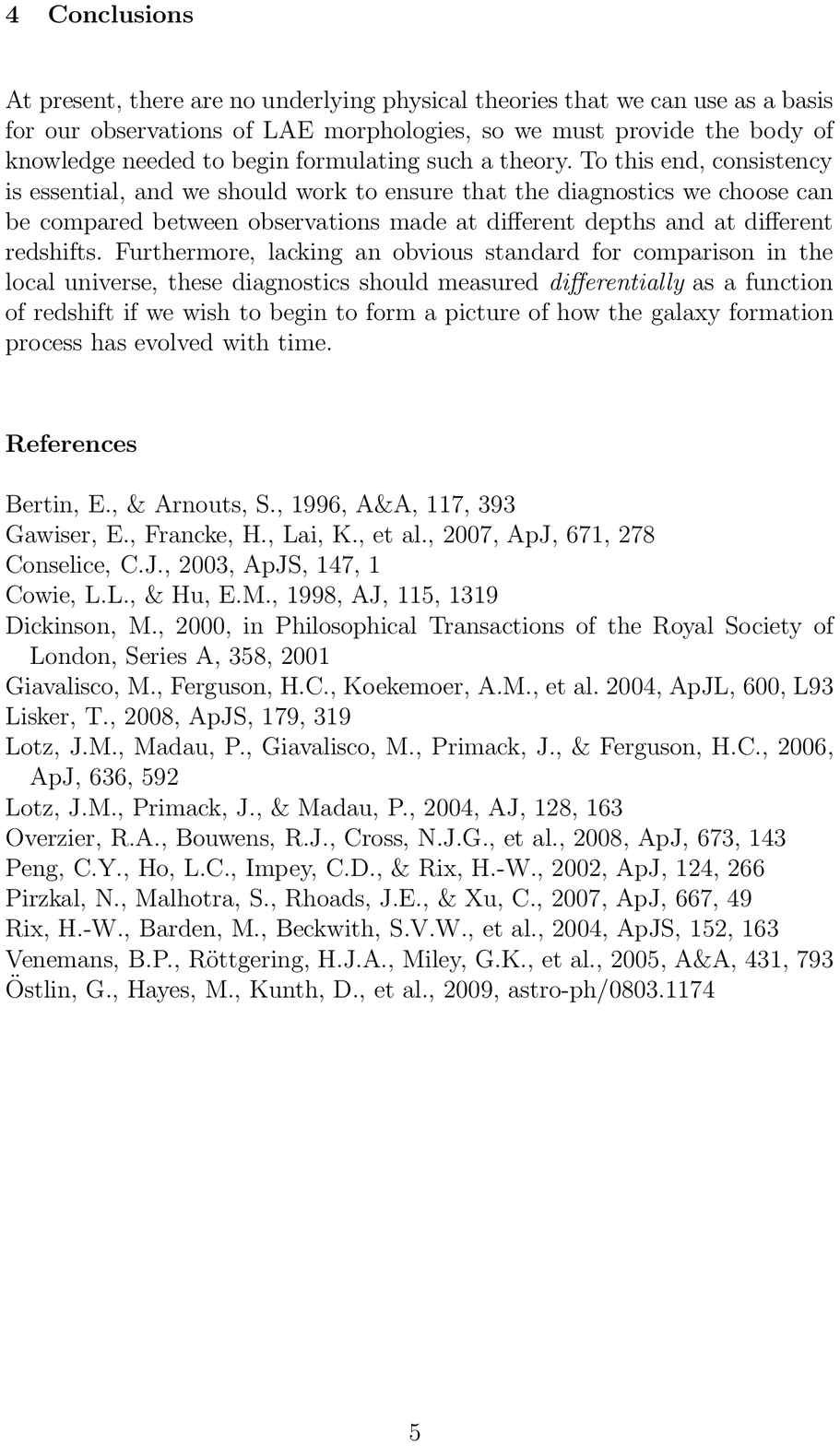,width=25cm,clip=}
\end{center}
\end{figure}
\newpage
\clearpage
\begin{figure}
\begin{center}
\epsfig{file=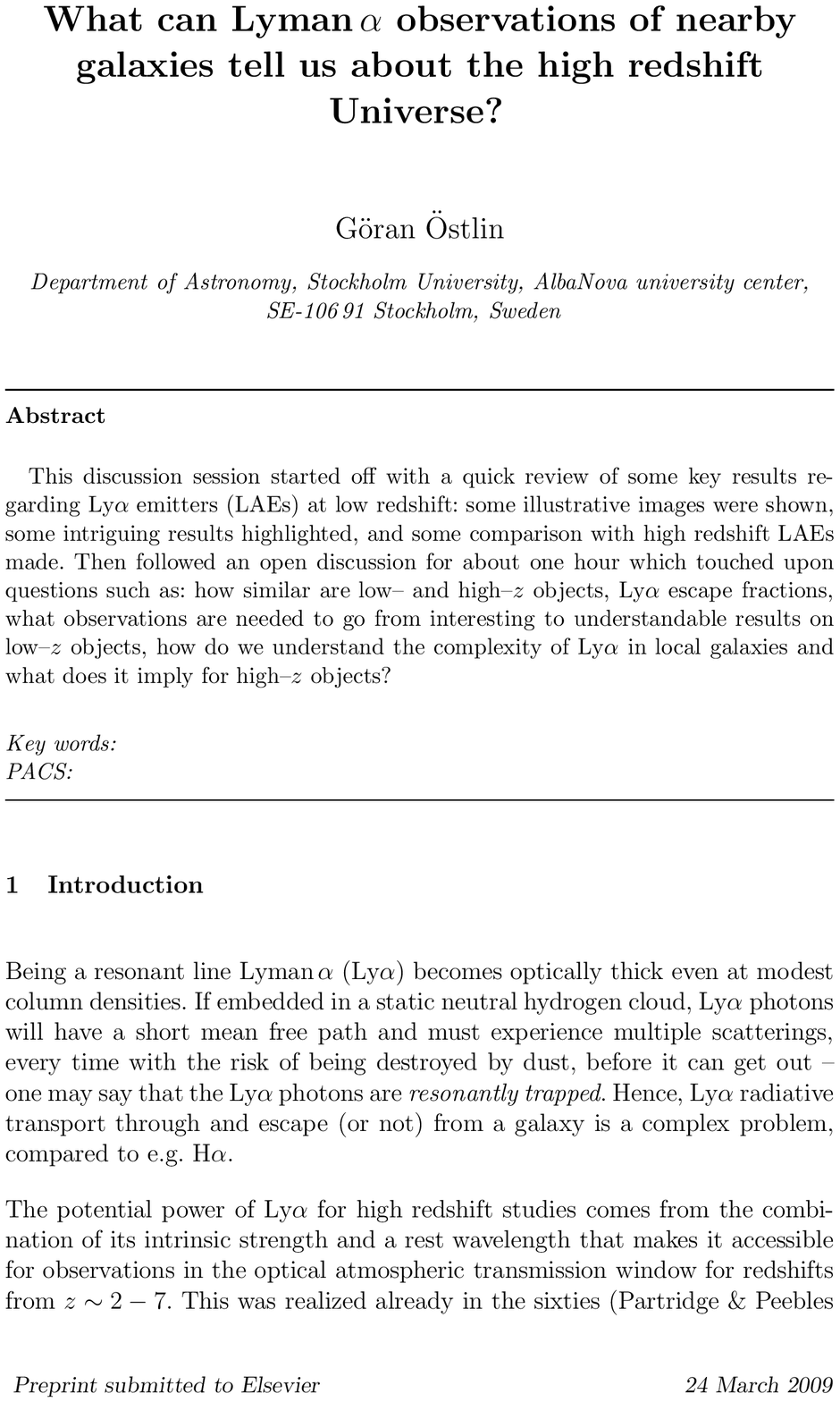,width=25cm,clip=}
\end{center}
\end{figure}
\newpage
\begin{figure}
\begin{center}
\epsfig{file=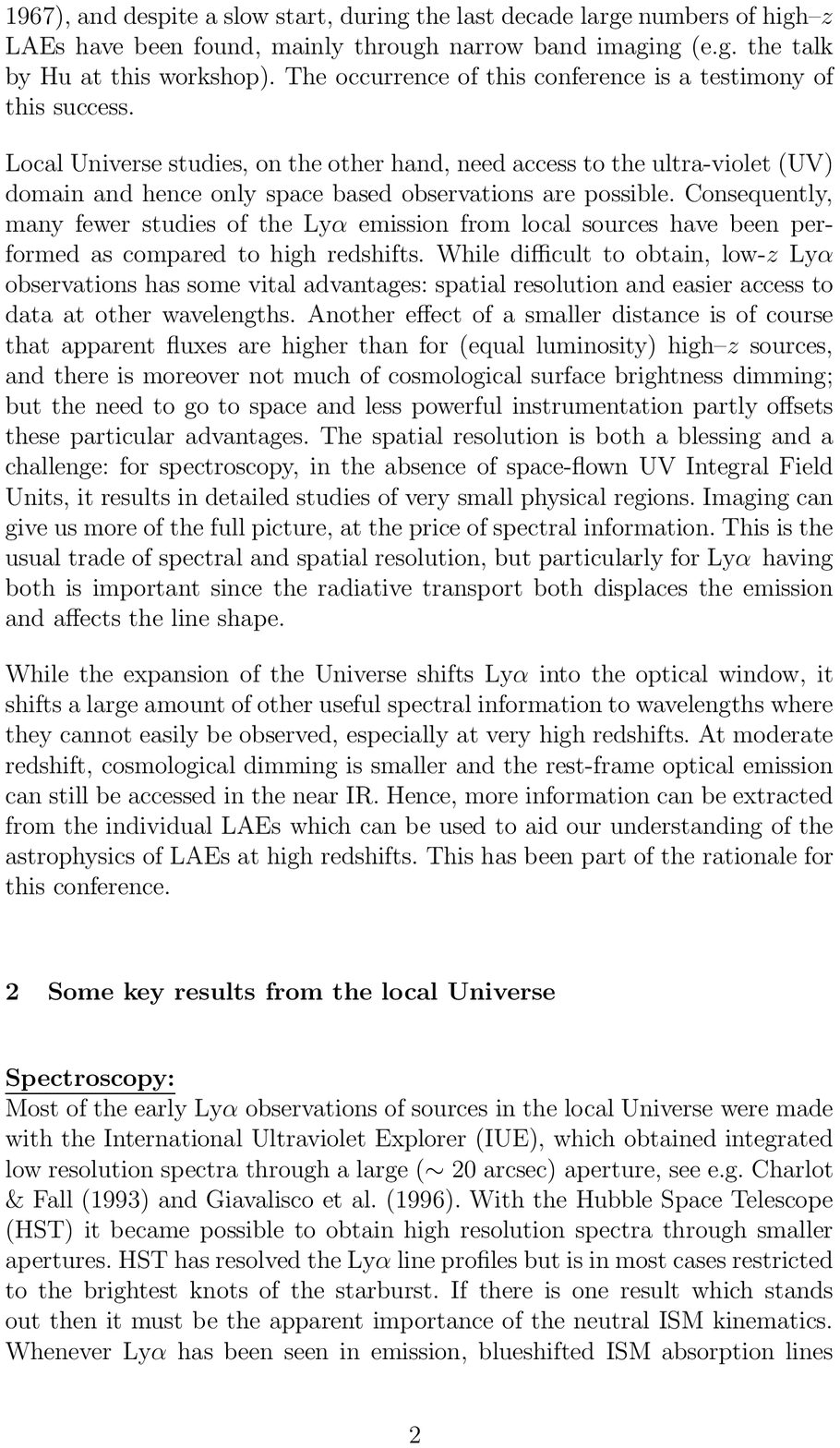,width=25cm,clip=}
\end{center}
\end{figure}
\newpage
\begin{figure}
\begin{center}
\epsfig{file=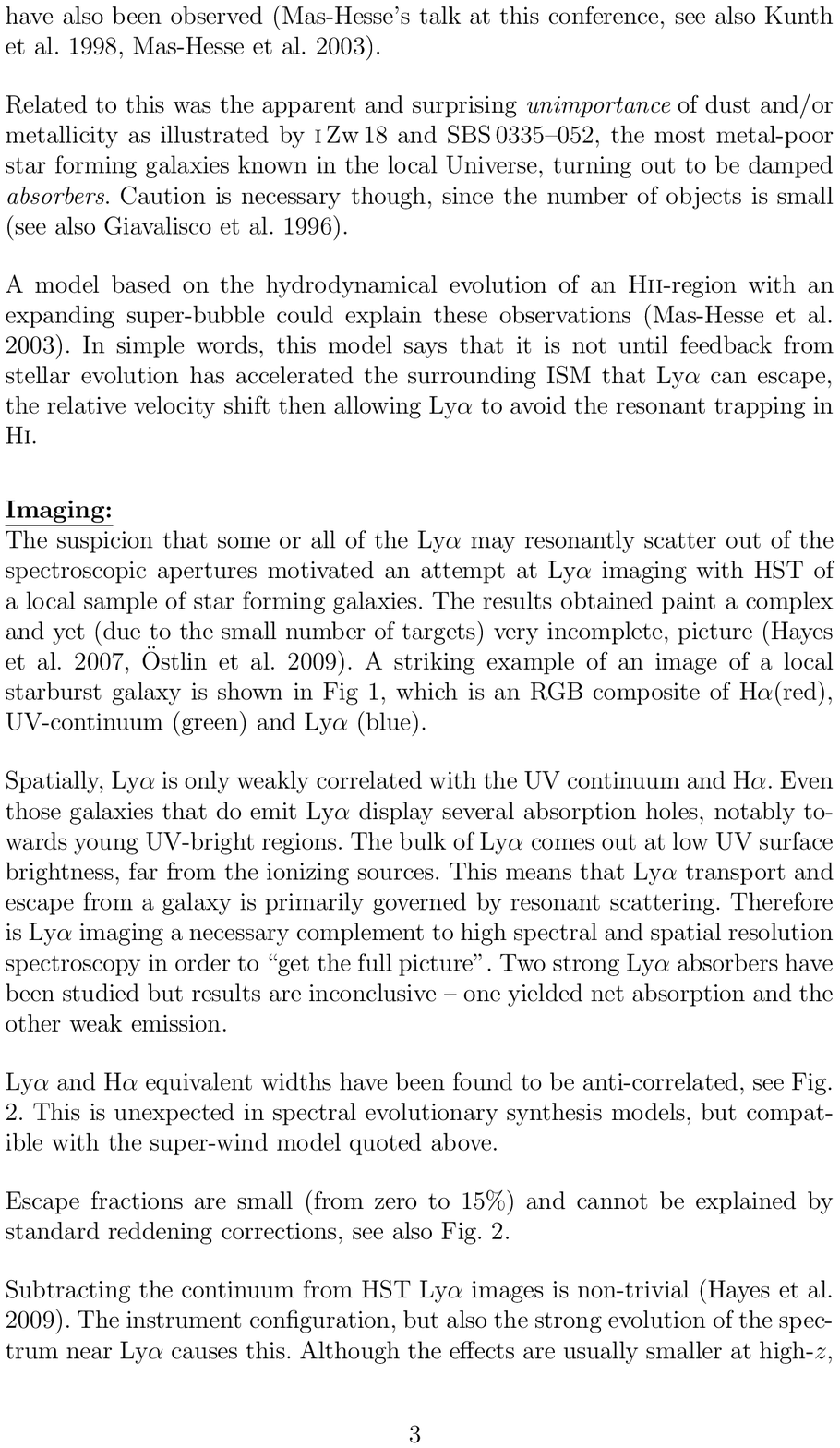,width=25cm,clip=}
\end{center}
\end{figure}
\newpage
\begin{figure}
\begin{center}
\epsfig{file=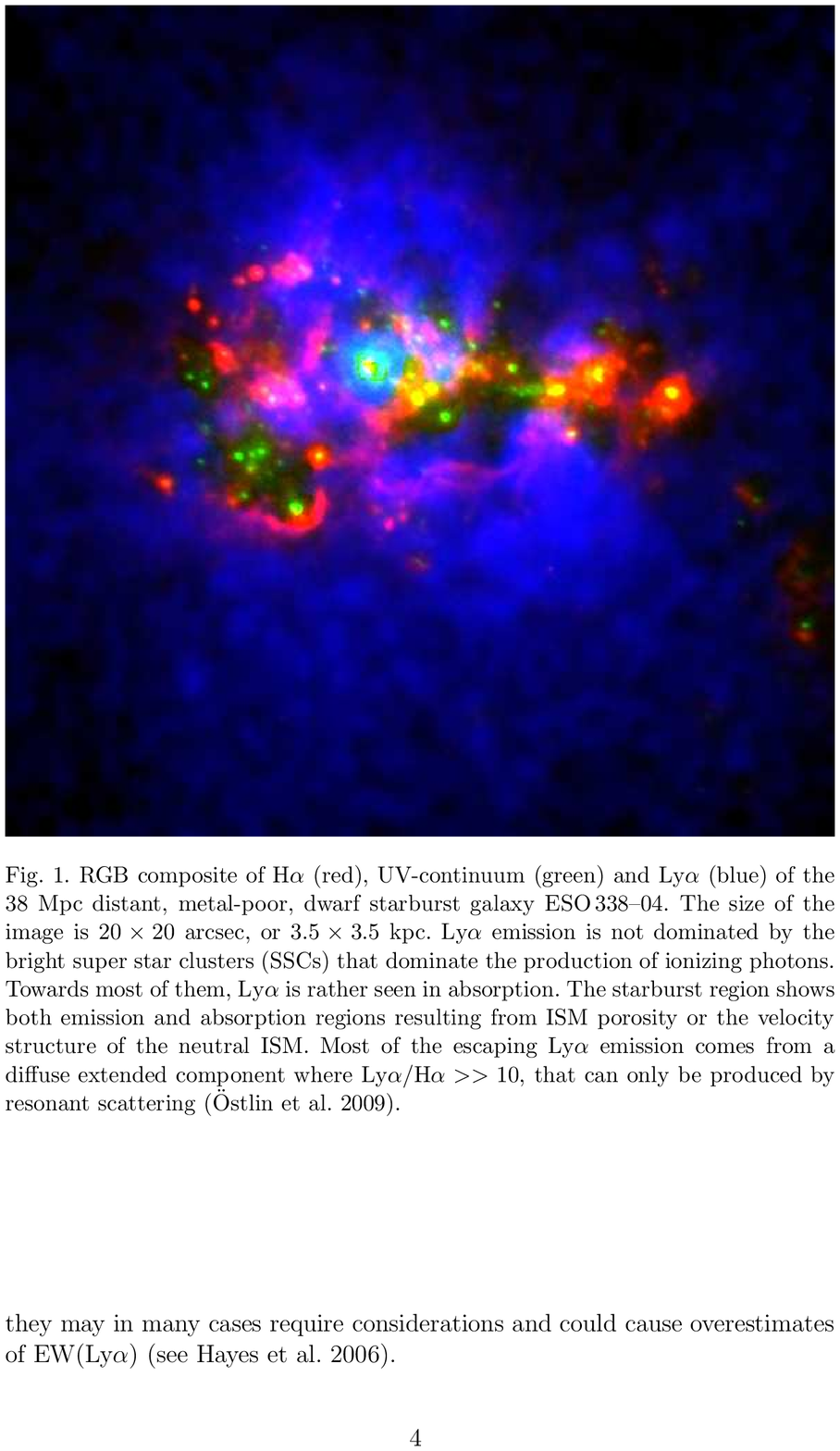,width=25cm,clip=}
\end{center}
\end{figure}
\newpage
\begin{figure}
\begin{center}
\epsfig{file=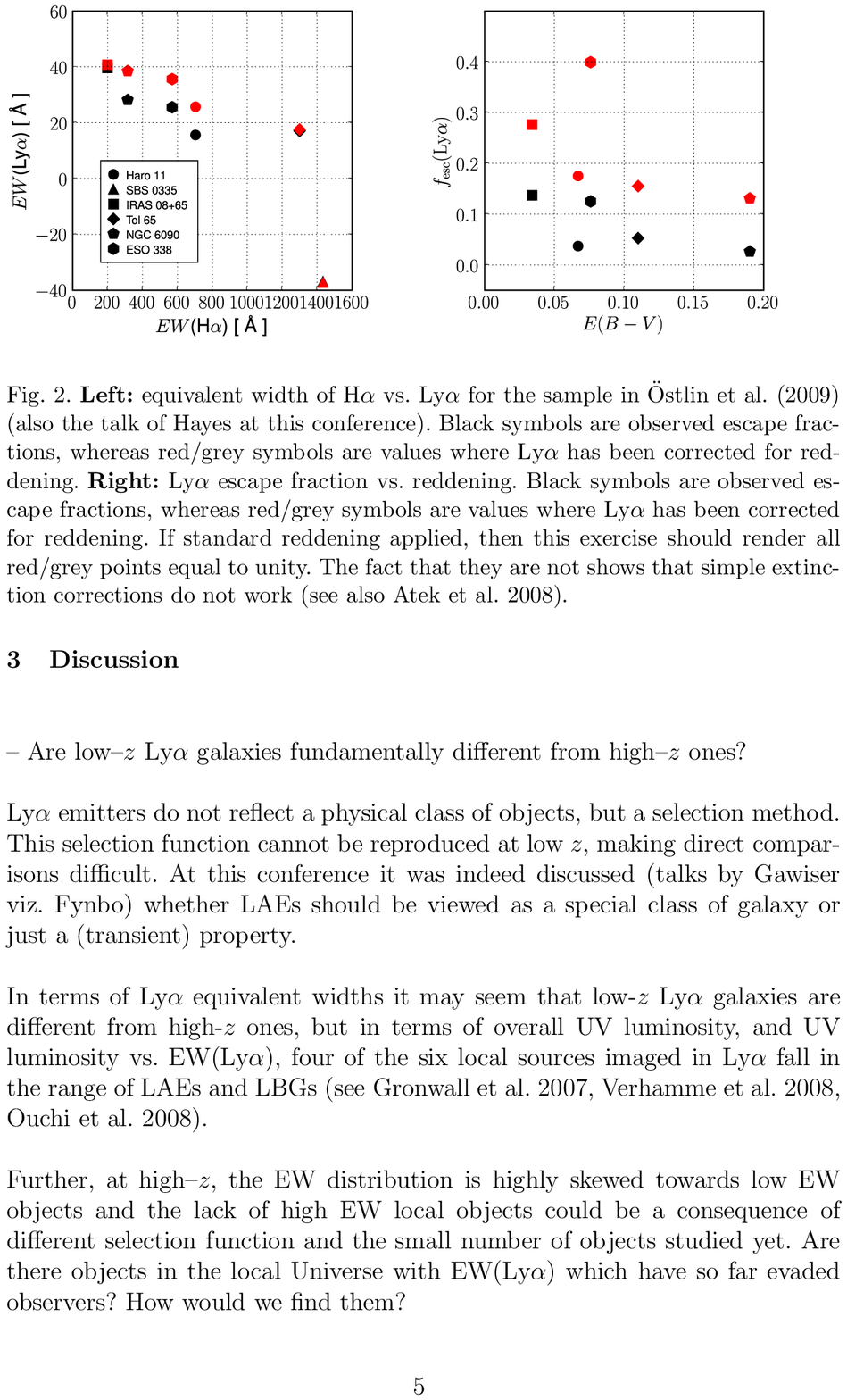,width=25cm,clip=}
\end{center}
\end{figure}
\newpage
\begin{figure}
\begin{center}
\epsfig{file=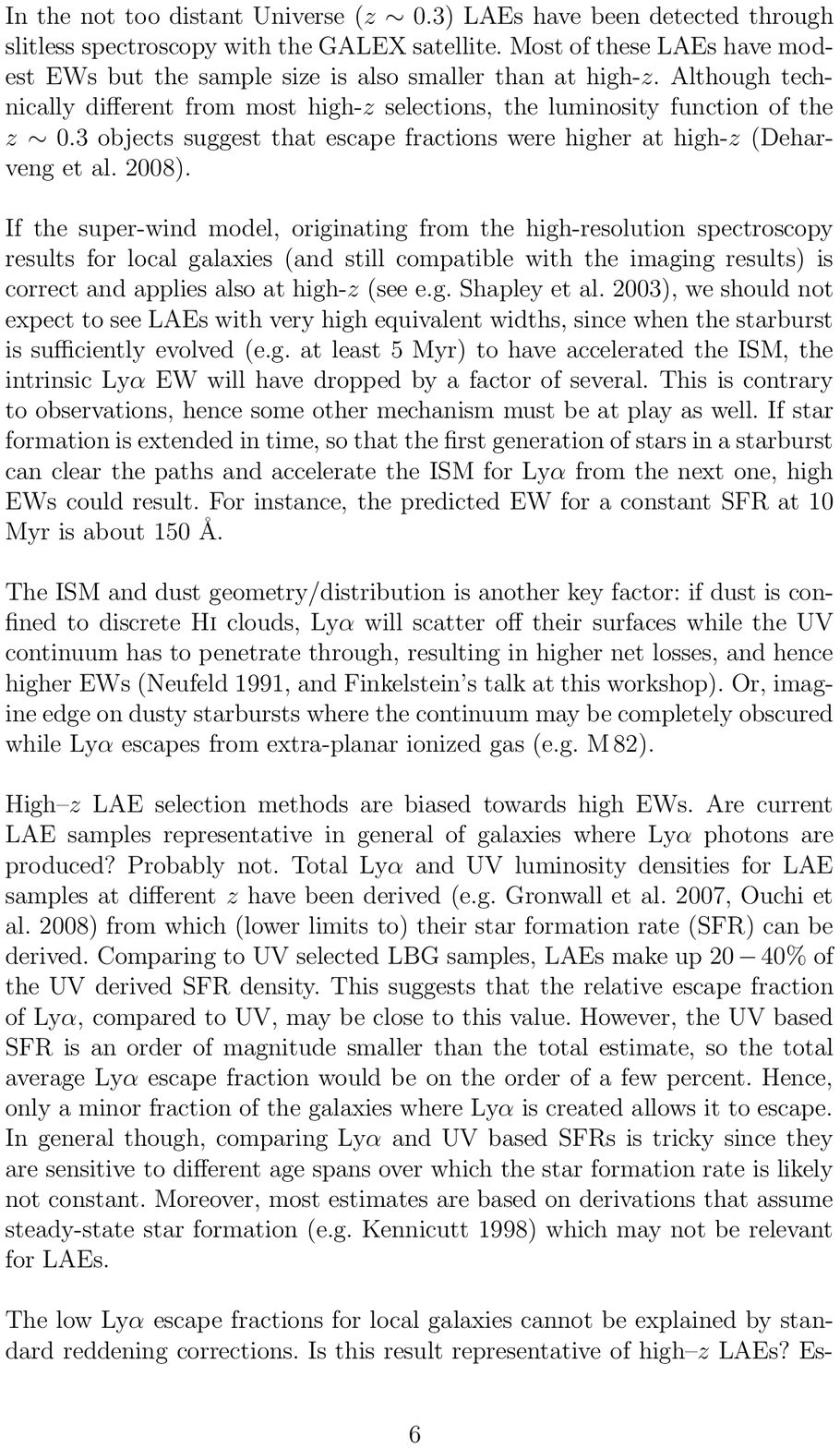,width=25cm,clip=}
\end{center}
\end{figure}
\newpage
\begin{figure}
\begin{center}
\epsfig{file=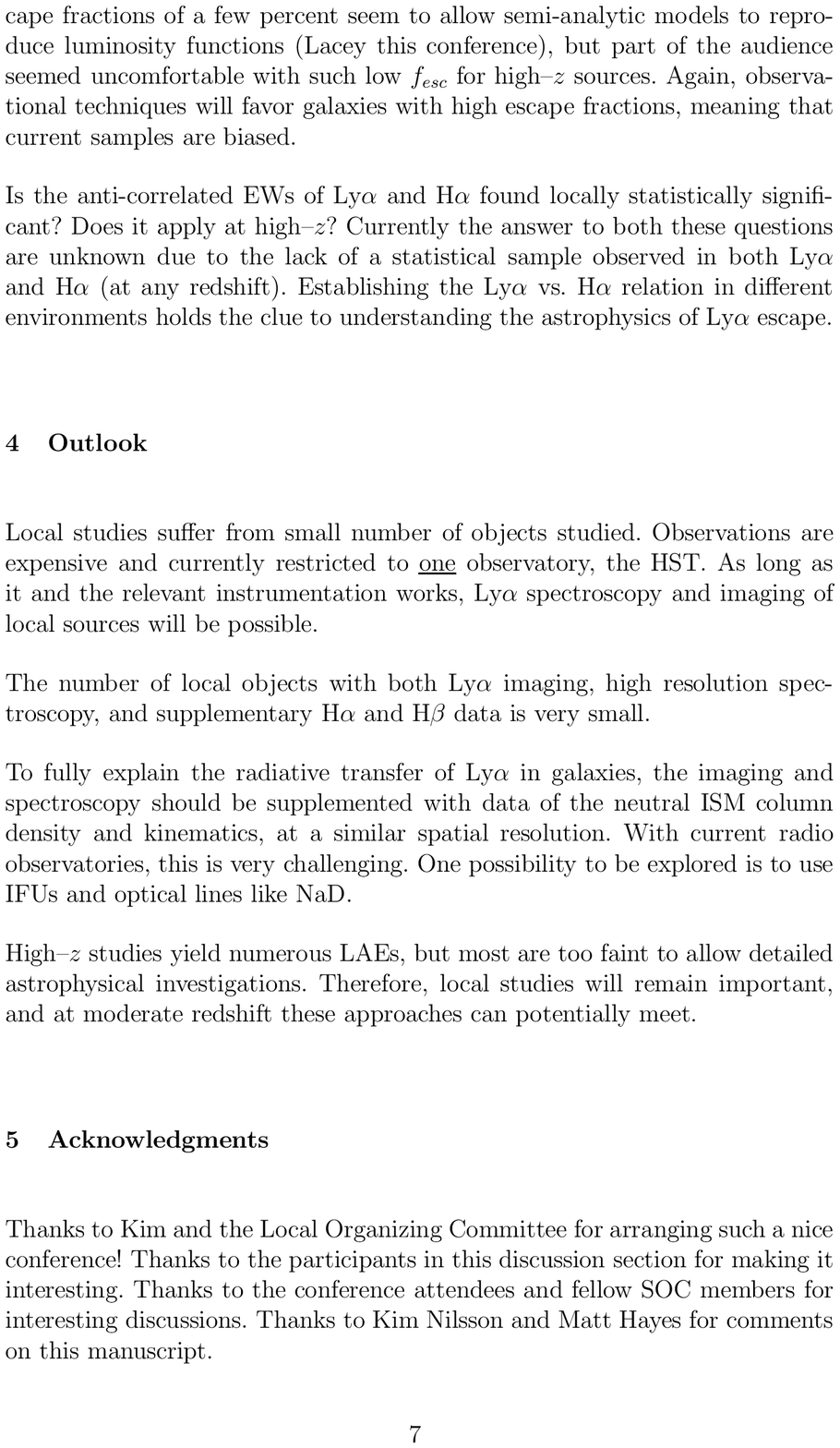,width=25cm,clip=}
\end{center}
\end{figure}
\newpage
\begin{figure}
\begin{center}
\epsfig{file=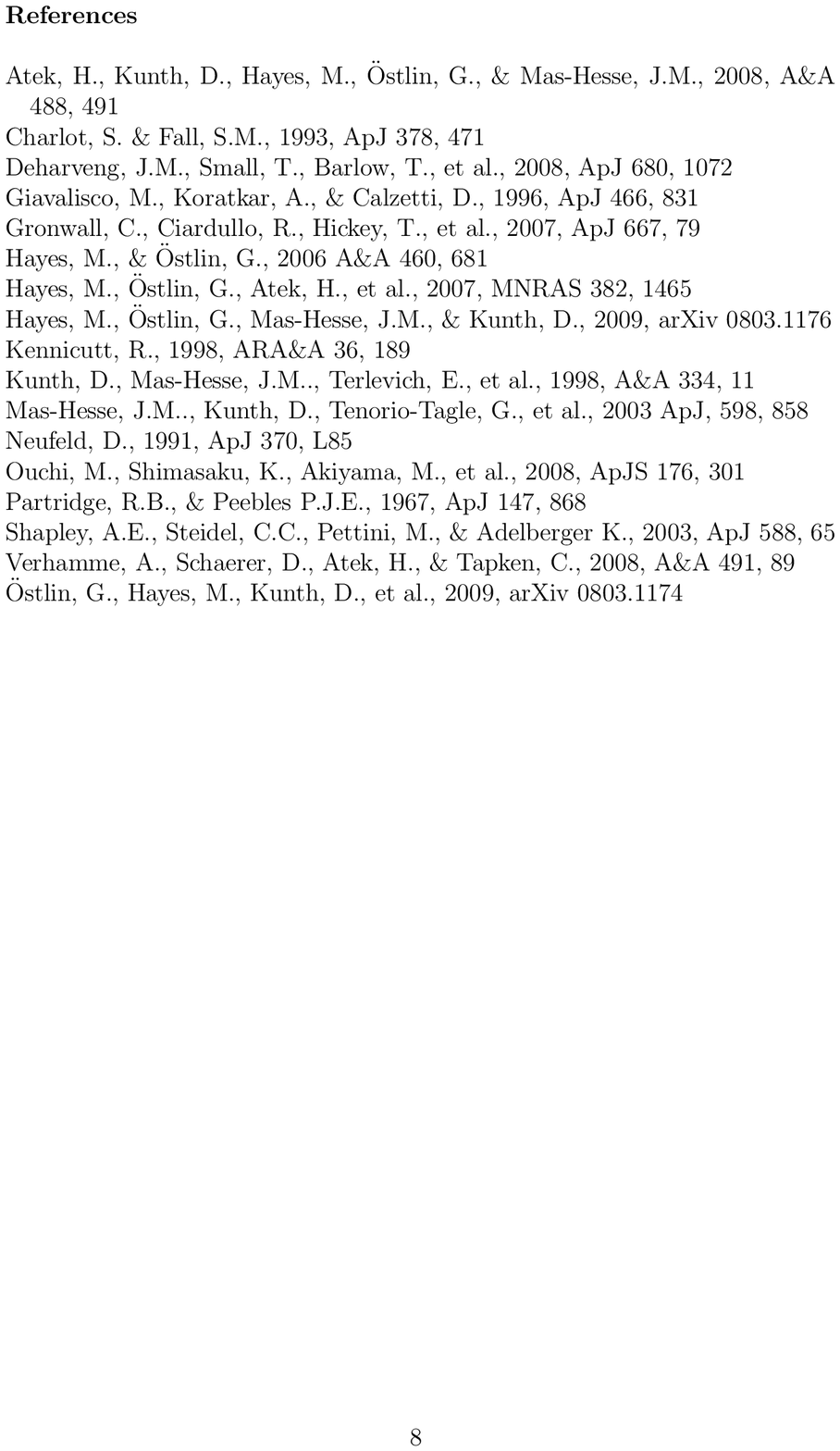,width=25cm,clip=}
\end{center}
\end{figure}
\newpage
\begin{figure}
\begin{center}
\epsfig{file=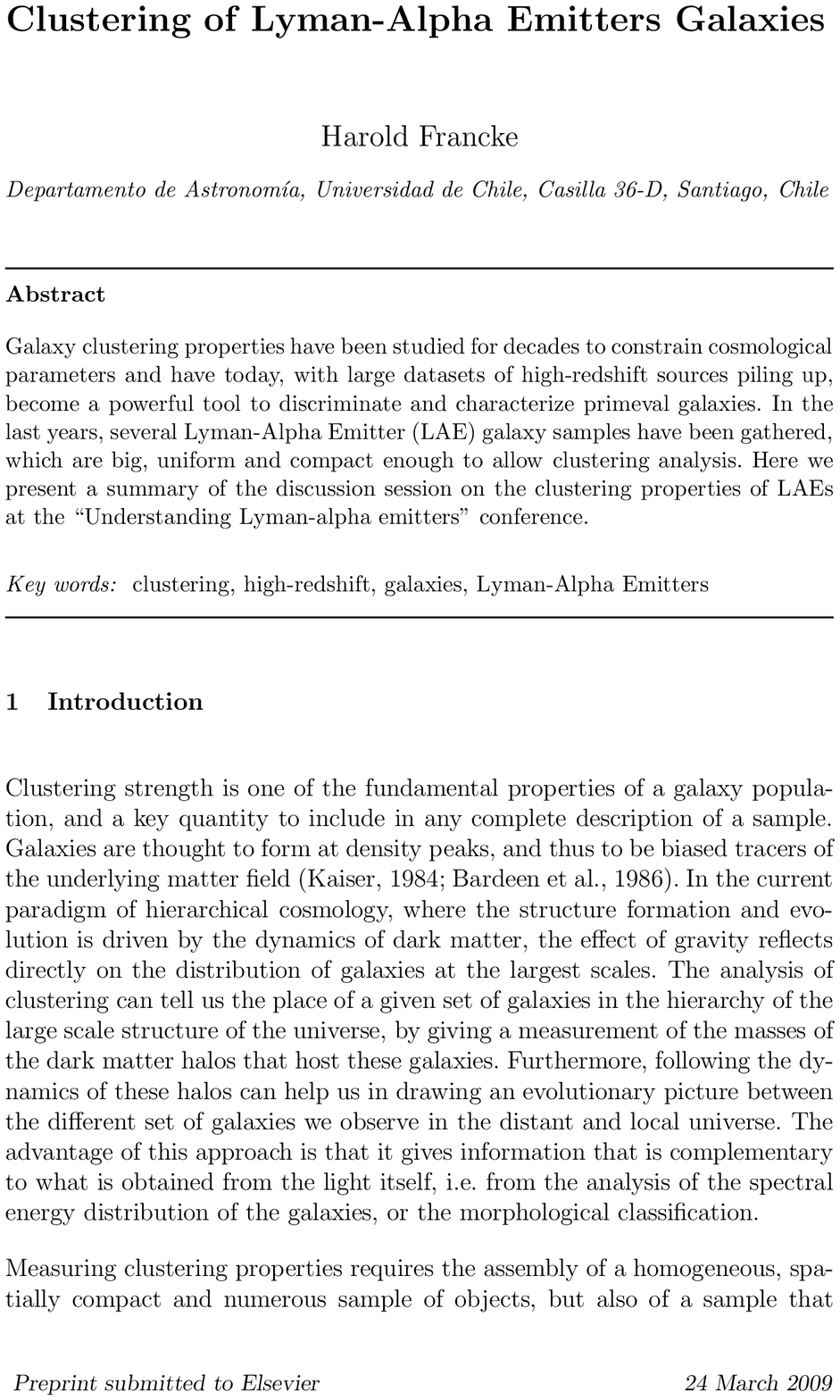,width=25cm,clip=}
\end{center}
\end{figure}
\newpage
\begin{figure}
\begin{center}
\epsfig{file=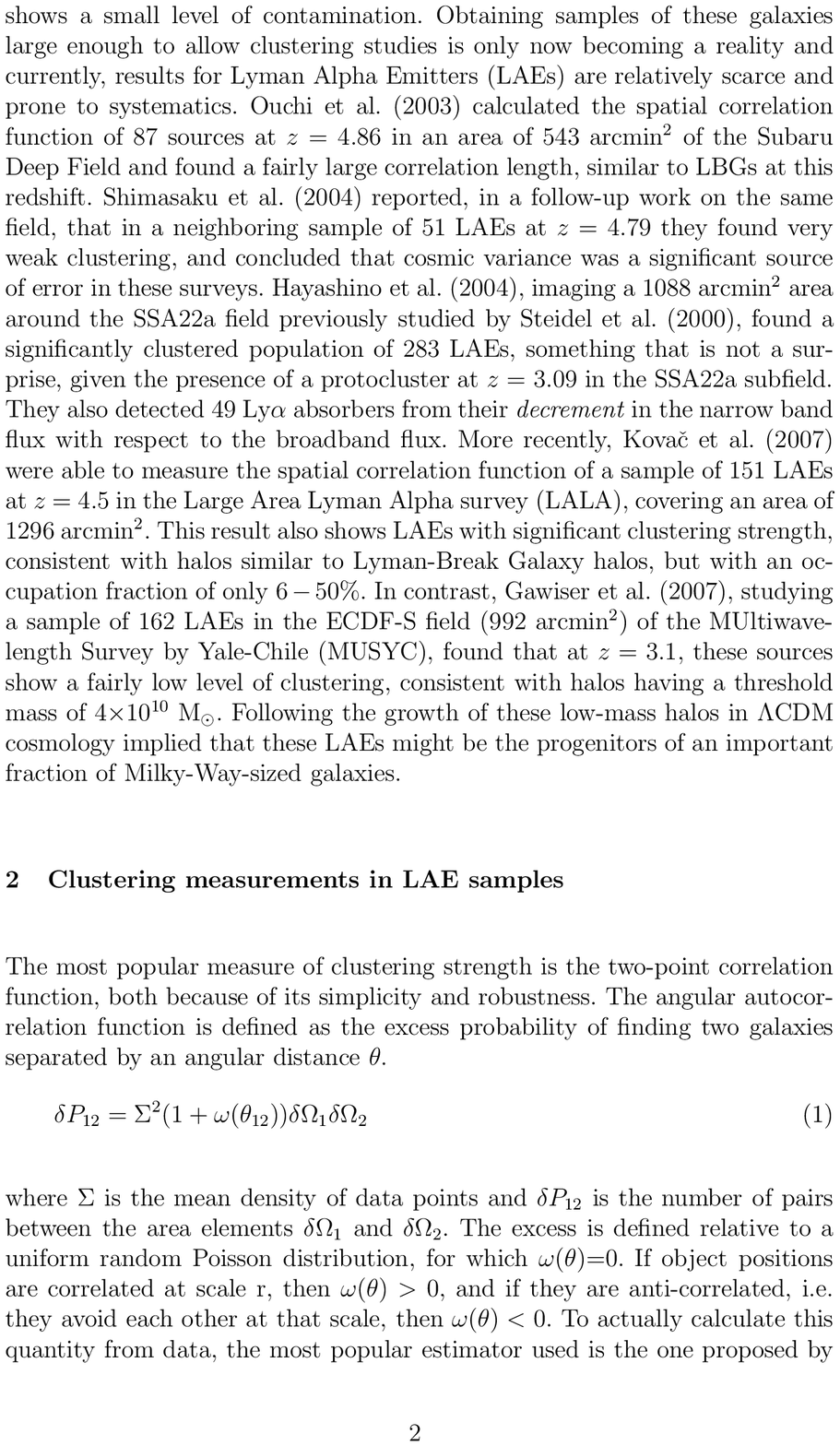,width=25cm,clip=}
\end{center}
\end{figure}
\newpage
\begin{figure}
\begin{center}
\epsfig{file=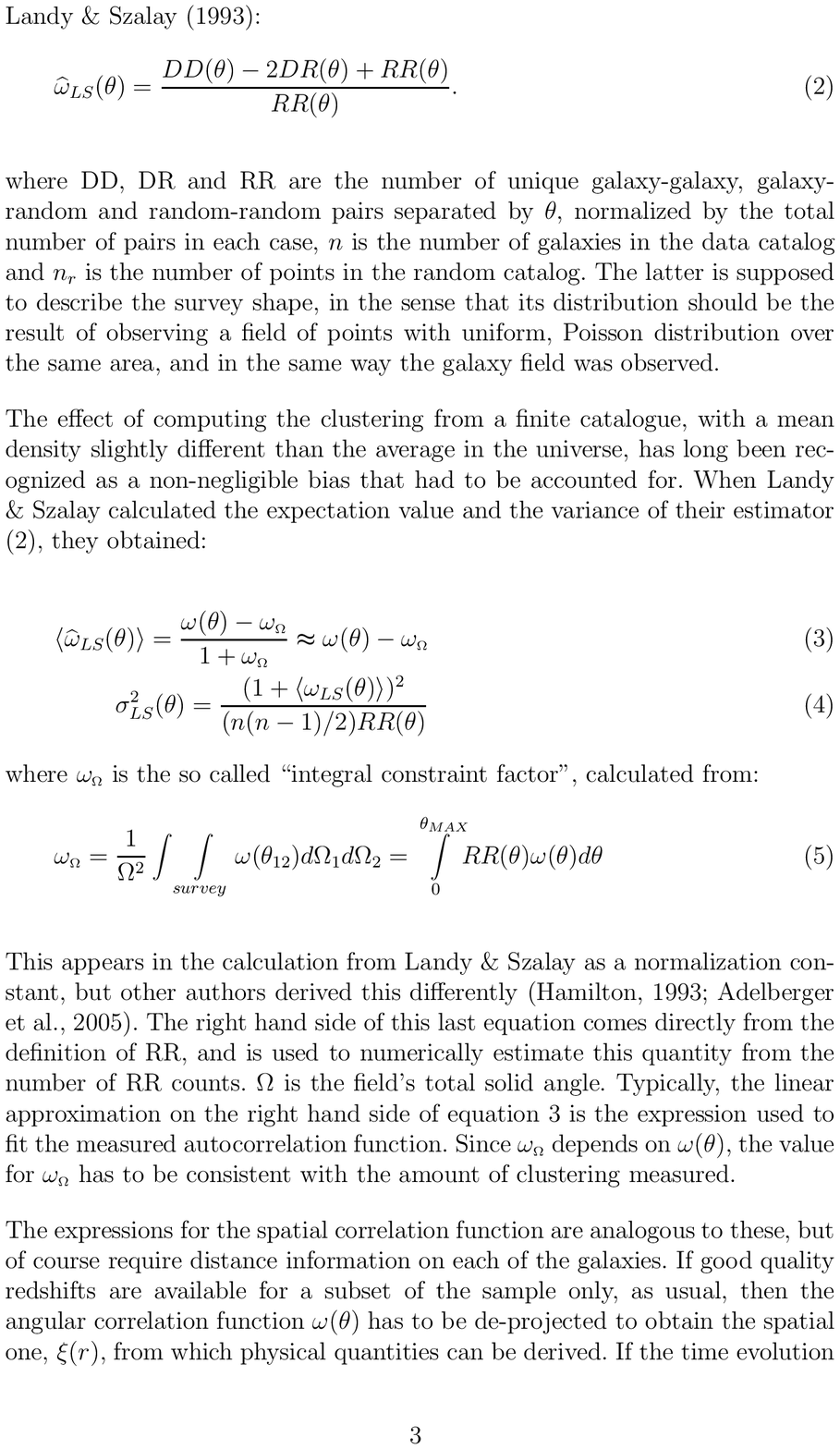,width=25cm,clip=}
\end{center}
\end{figure}
\newpage
\begin{figure}
\begin{center}
\epsfig{file=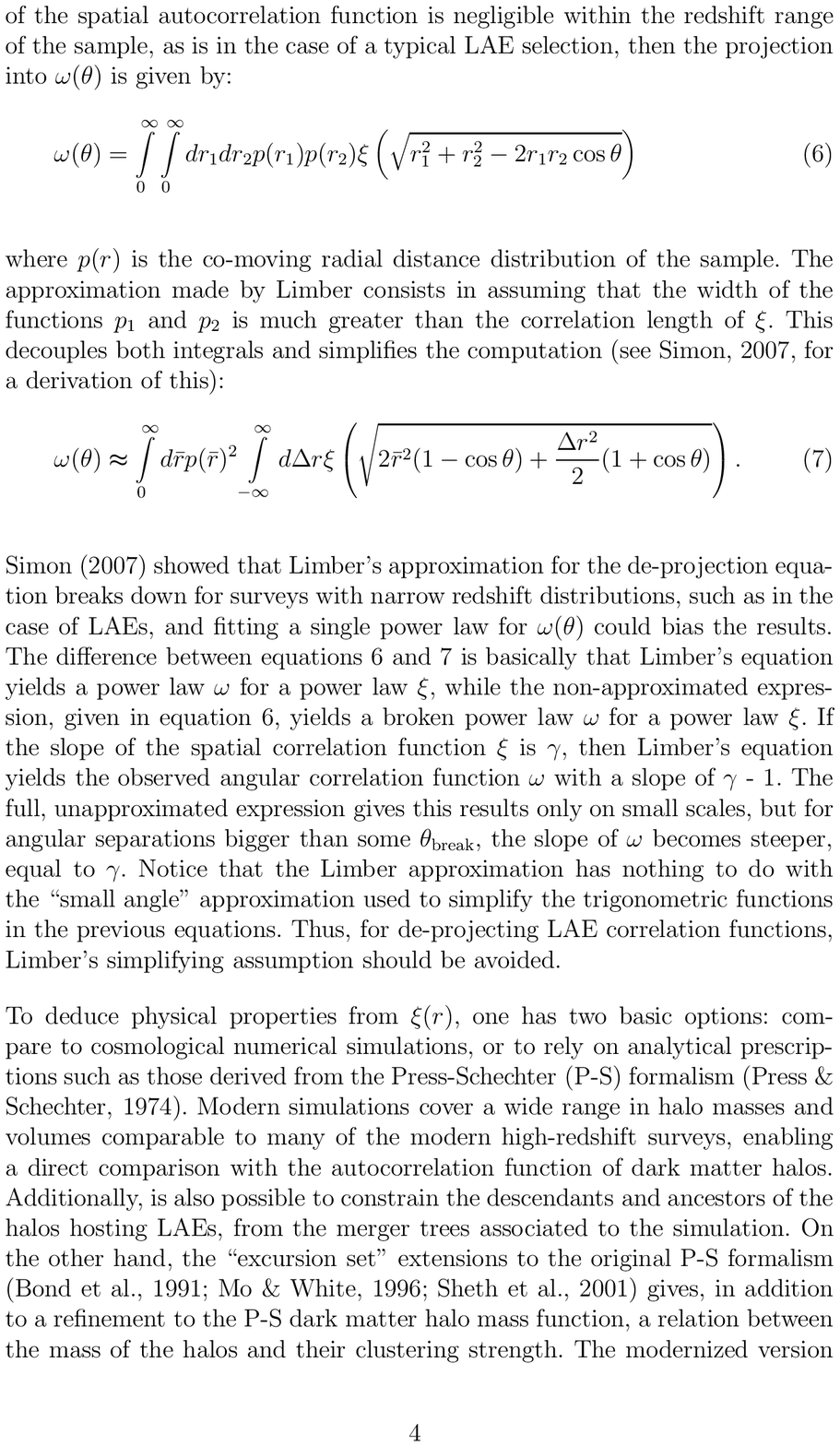,width=25cm,clip=}
\end{center}
\end{figure}
\newpage
\begin{figure}
\begin{center}
\epsfig{file=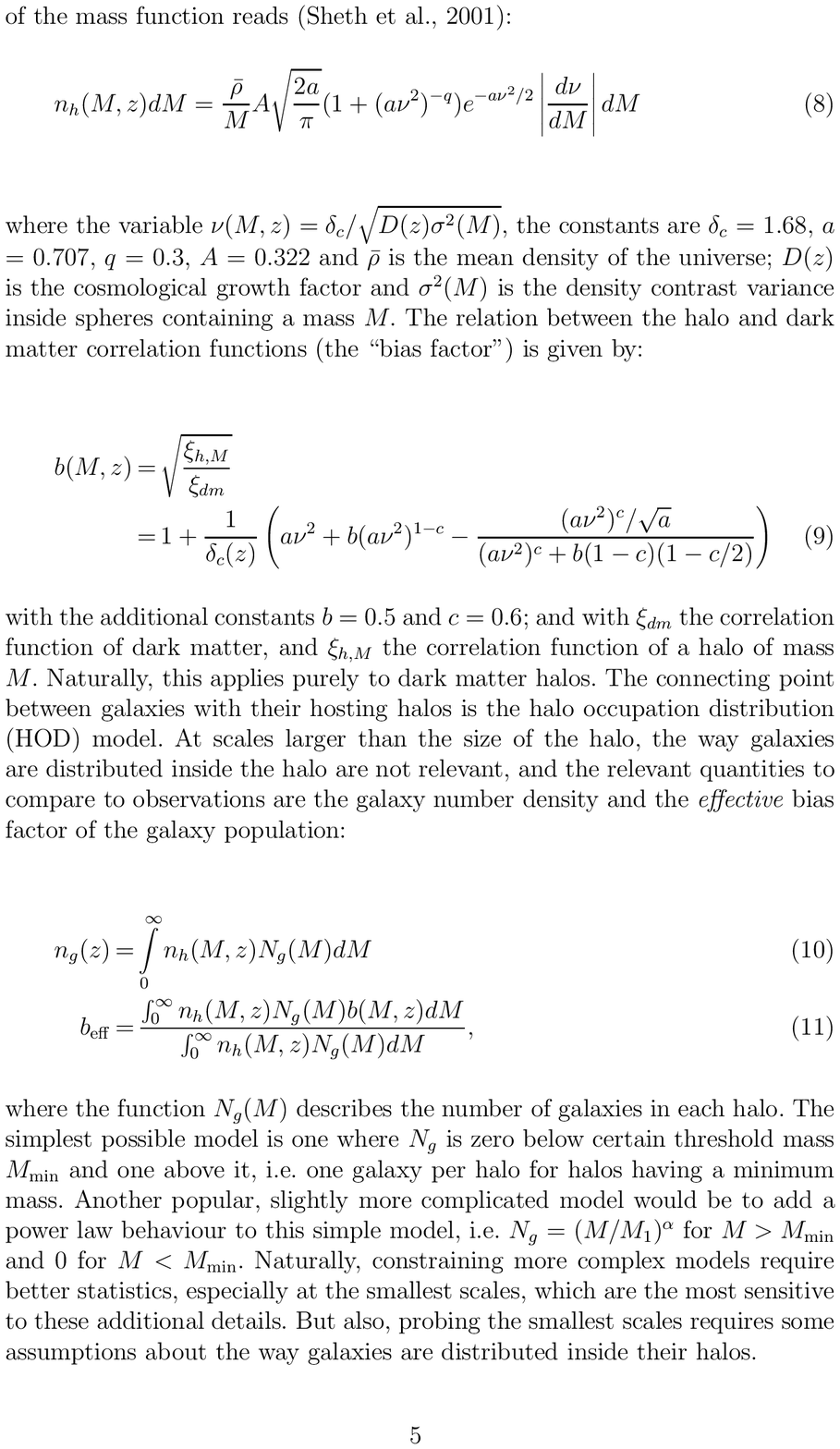,width=25cm,clip=}
\end{center}
\end{figure}
\newpage
\begin{figure}
\begin{center}
\epsfig{file=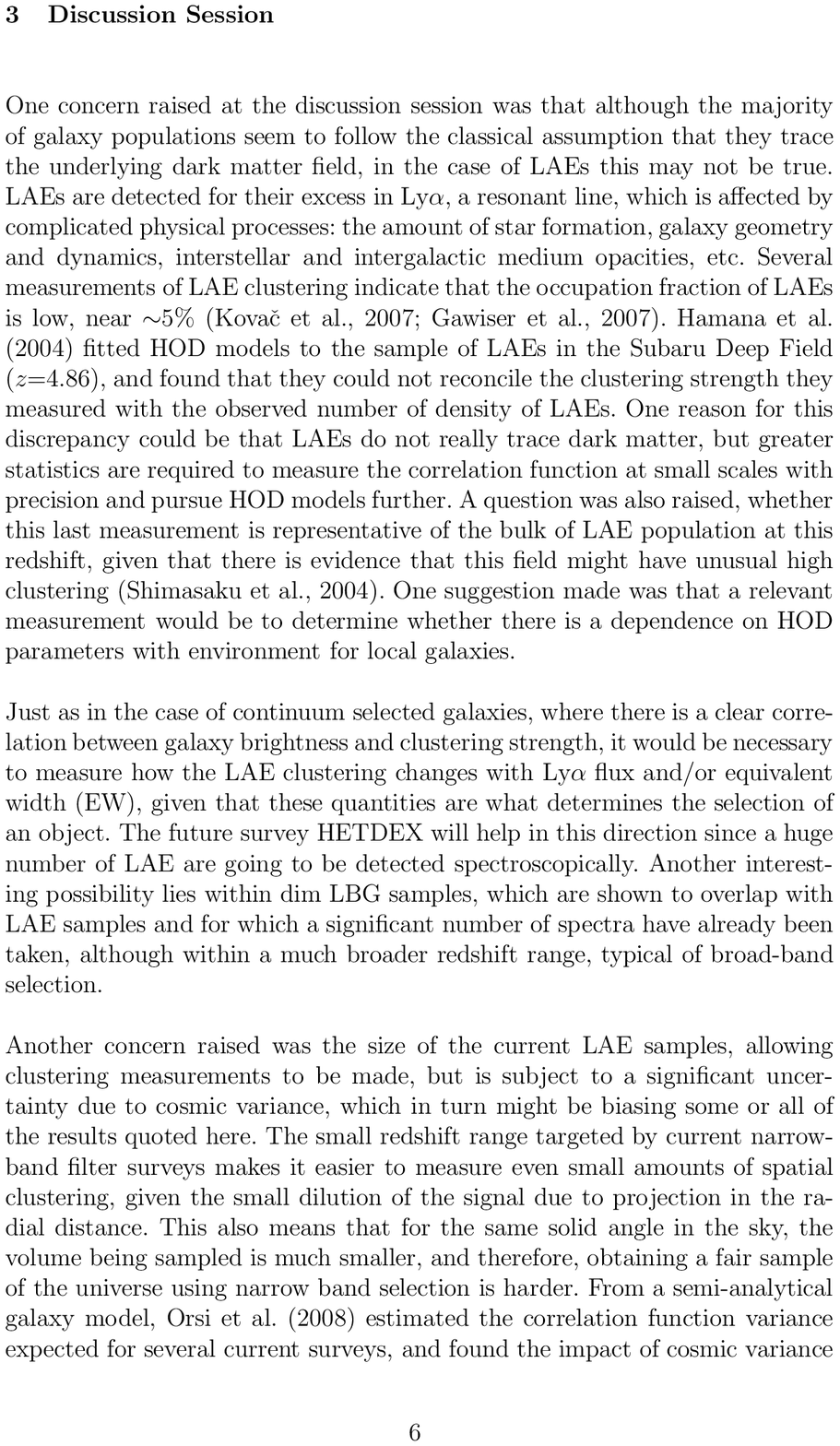,width=25cm,clip=}
\end{center}
\end{figure}
\newpage
\begin{figure}
\begin{center}
\epsfig{file=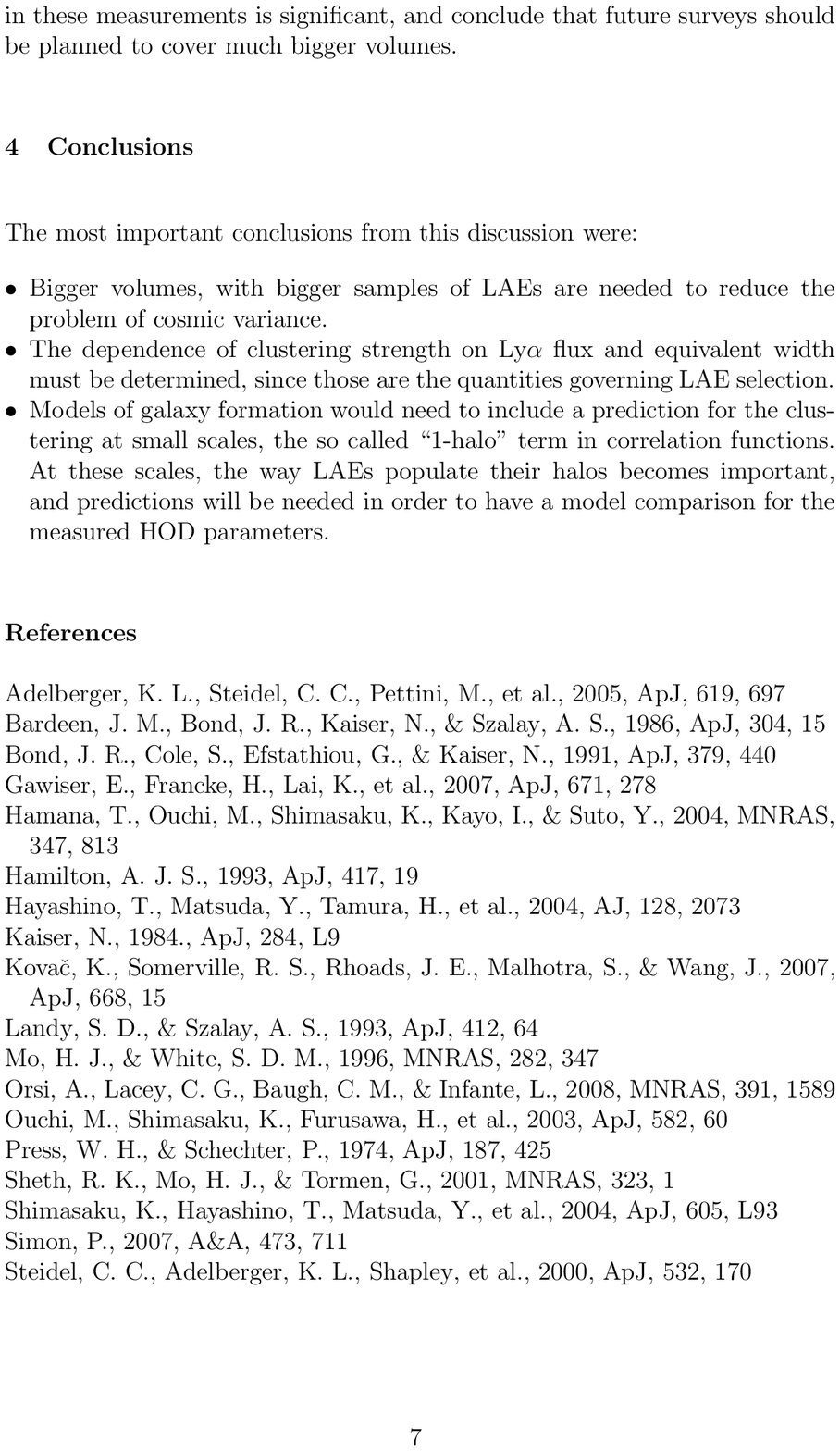,width=25cm,clip=}
\end{center}
\end{figure}
\newpage
\clearpage
\begin{figure}
\begin{center}
\epsfig{file=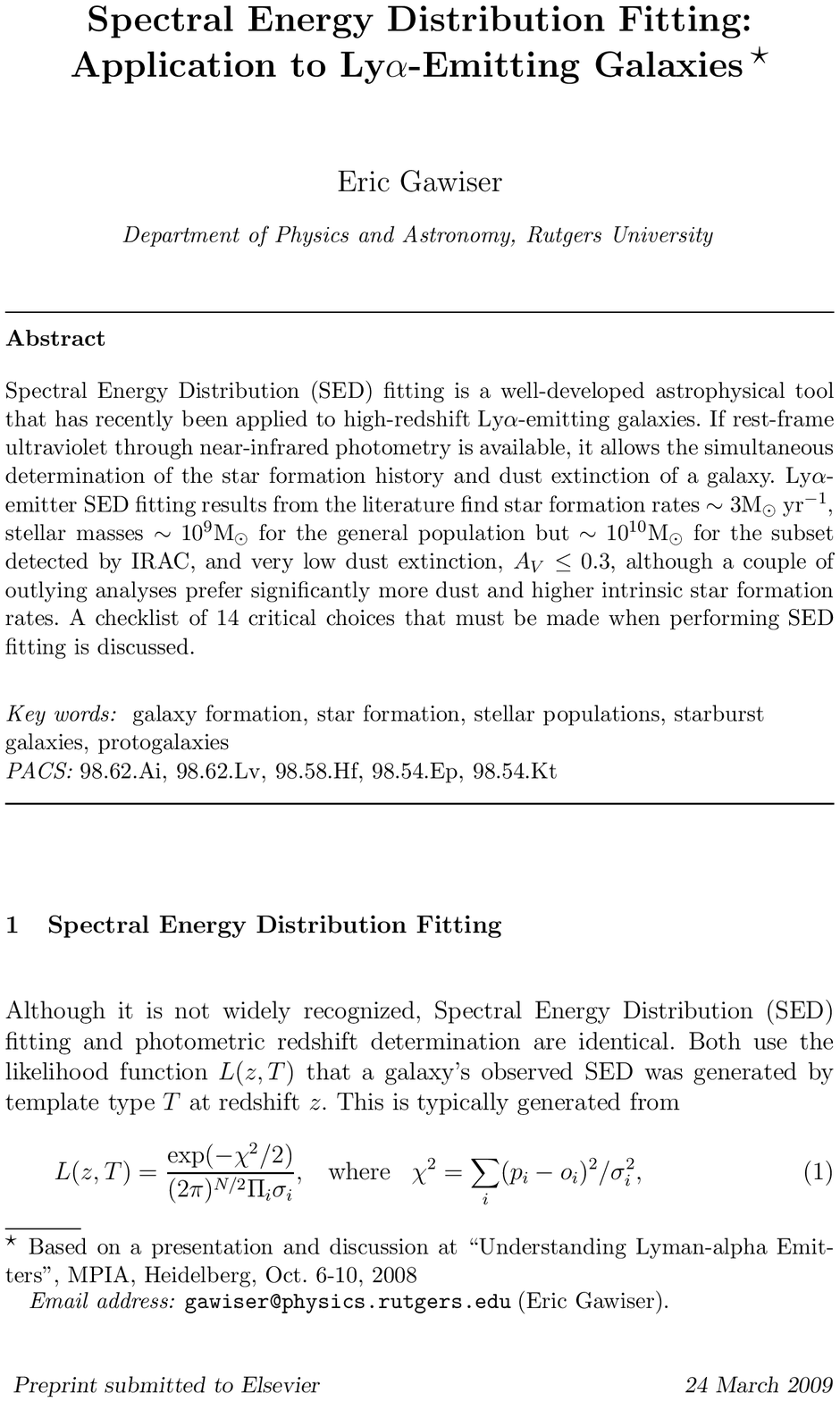,width=25cm,clip=}
\end{center}
\end{figure}
\newpage
\begin{figure}
\begin{center}
\epsfig{file=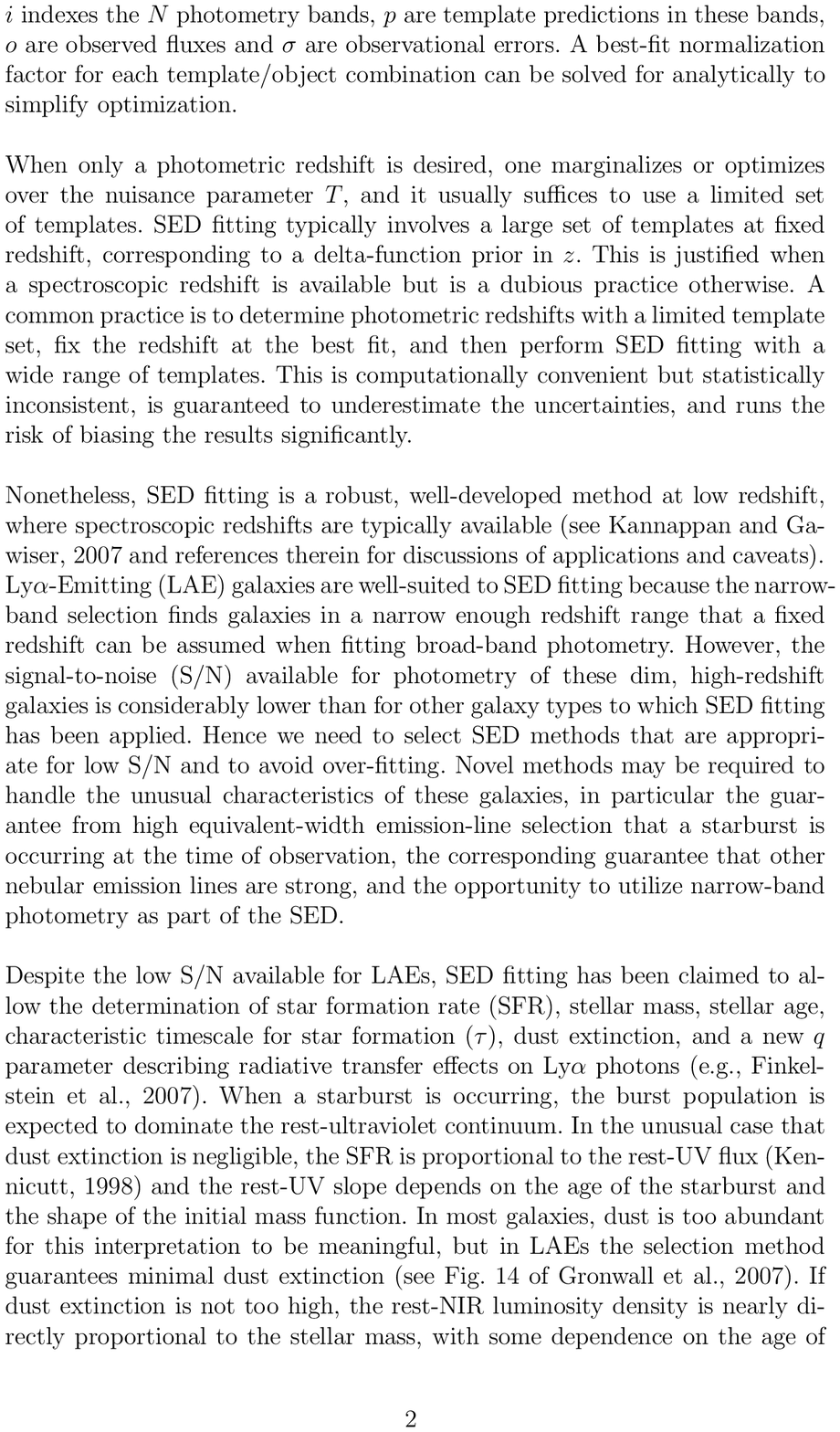,width=25cm,clip=}
\end{center}
\end{figure}
\newpage
\begin{figure}
\begin{center}
\epsfig{file=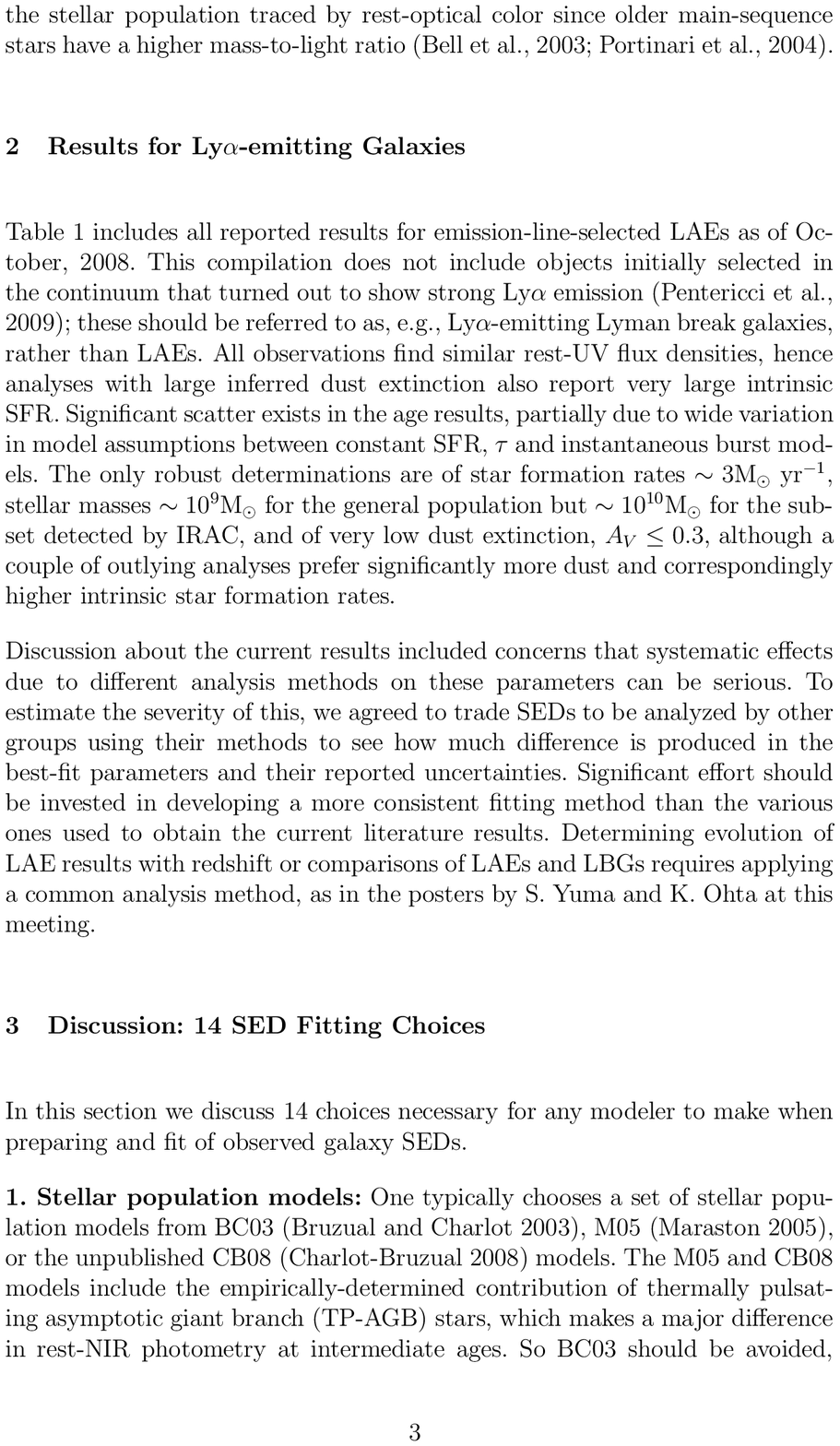,width=25cm,clip=}
\end{center}
\end{figure}
\newpage
\begin{figure}
\begin{center}
\epsfig{file=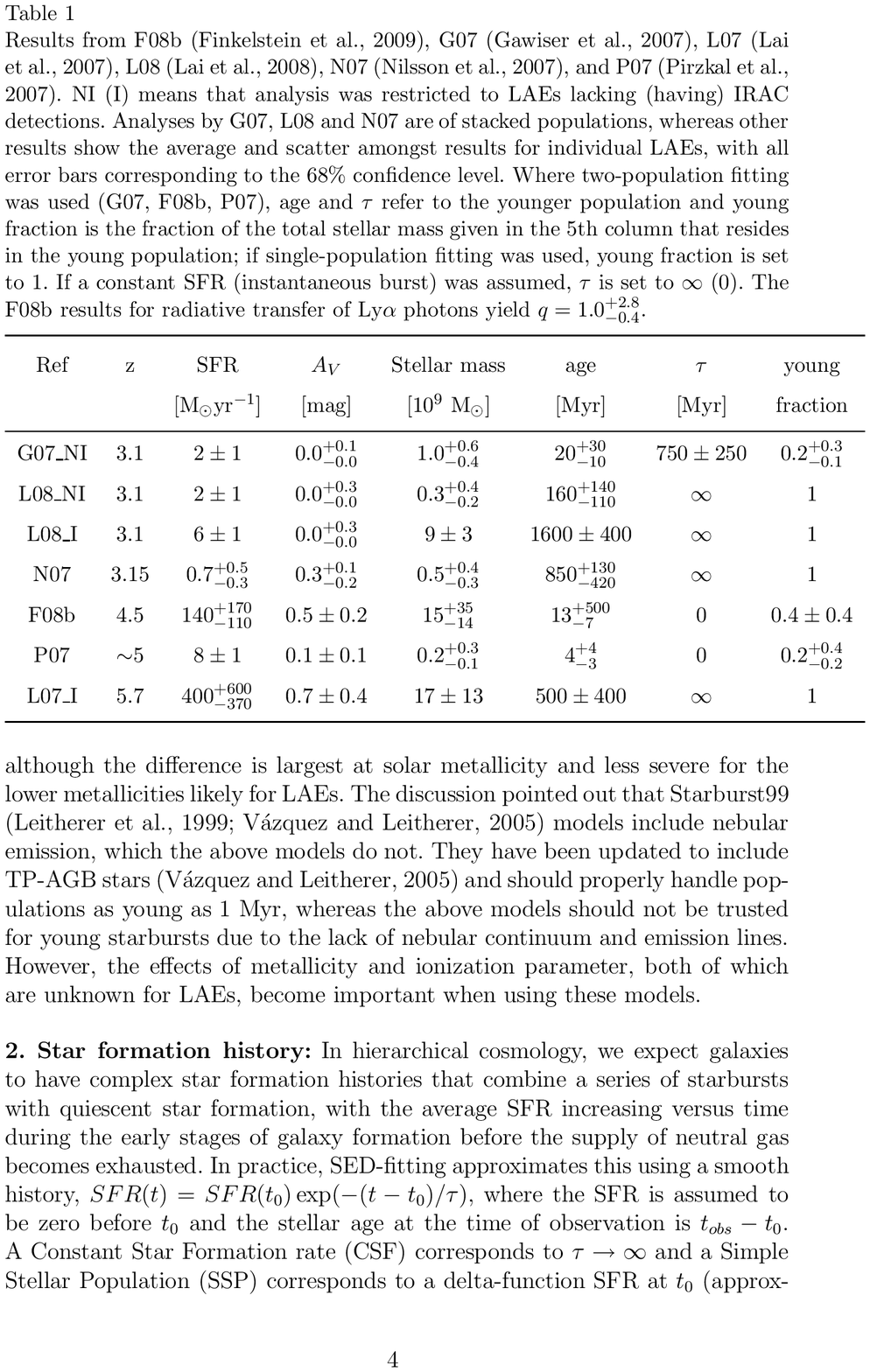,width=25cm,clip=}
\end{center}
\end{figure}
\newpage
\begin{figure}
\begin{center}
\epsfig{file=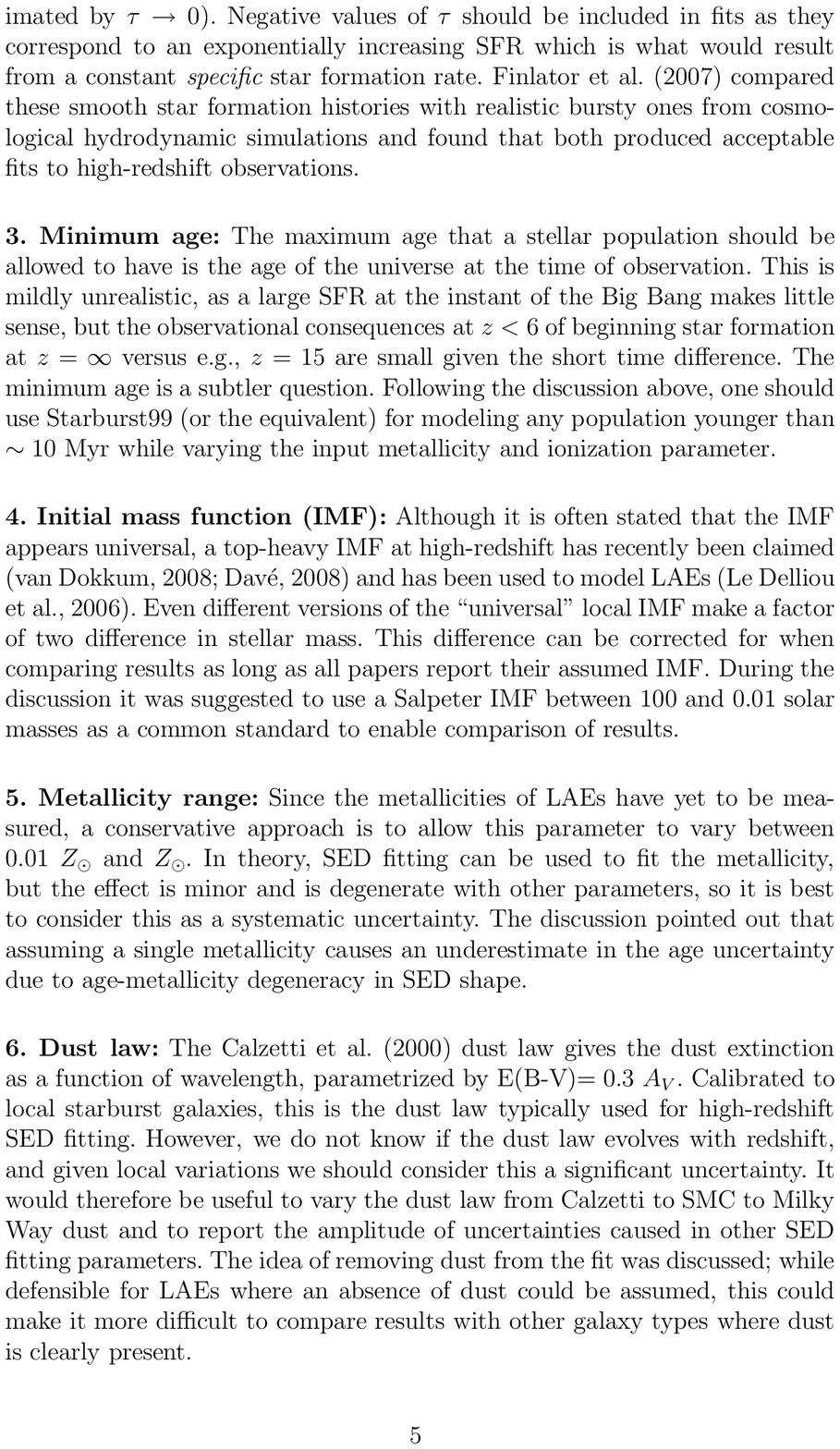,width=25cm,clip=}
\end{center}
\end{figure}
\newpage
\begin{figure}
\begin{center}
\epsfig{file=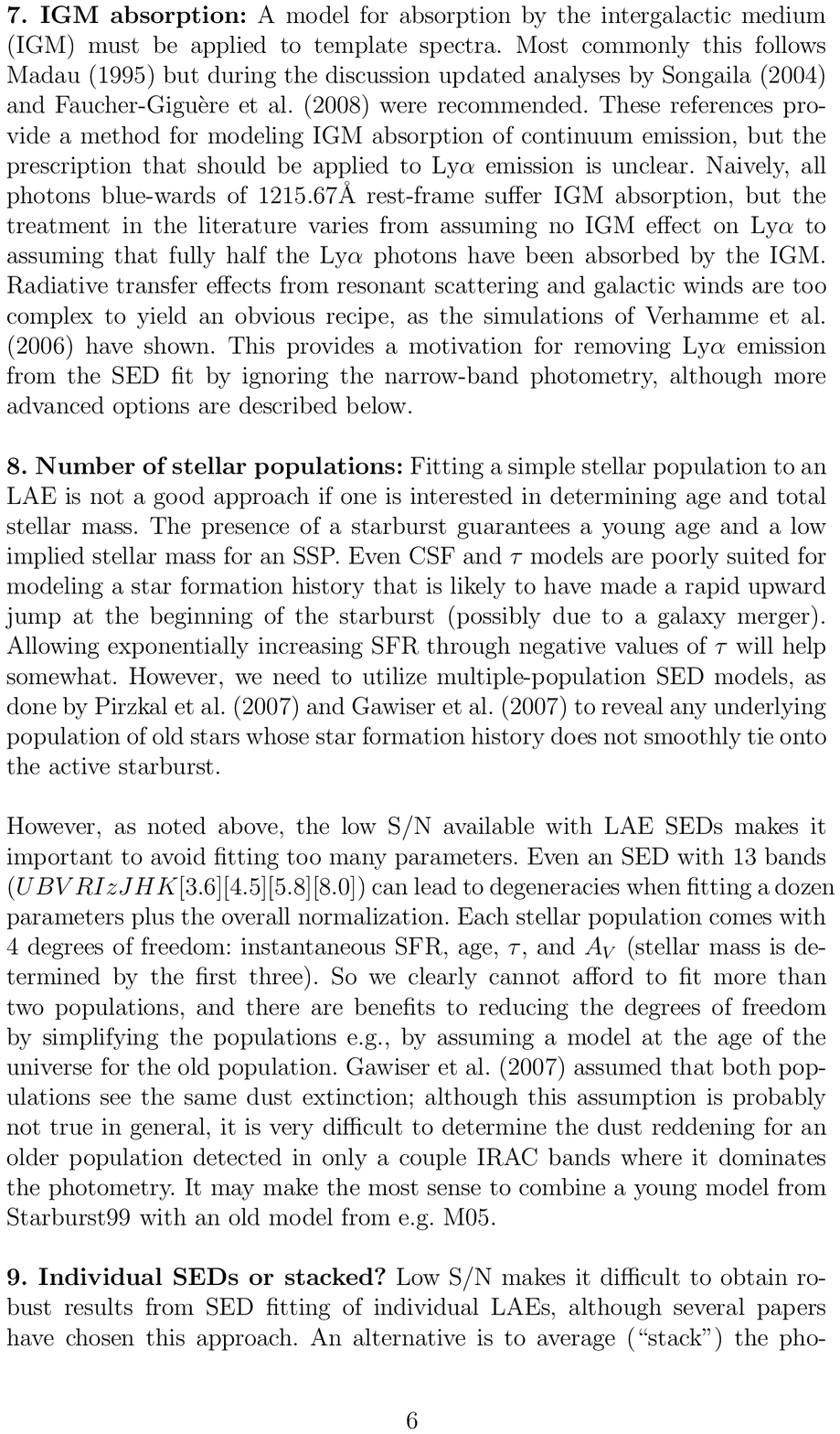,width=25cm,clip=}
\end{center}
\end{figure}
\newpage
\begin{figure}
\begin{center}
\epsfig{file=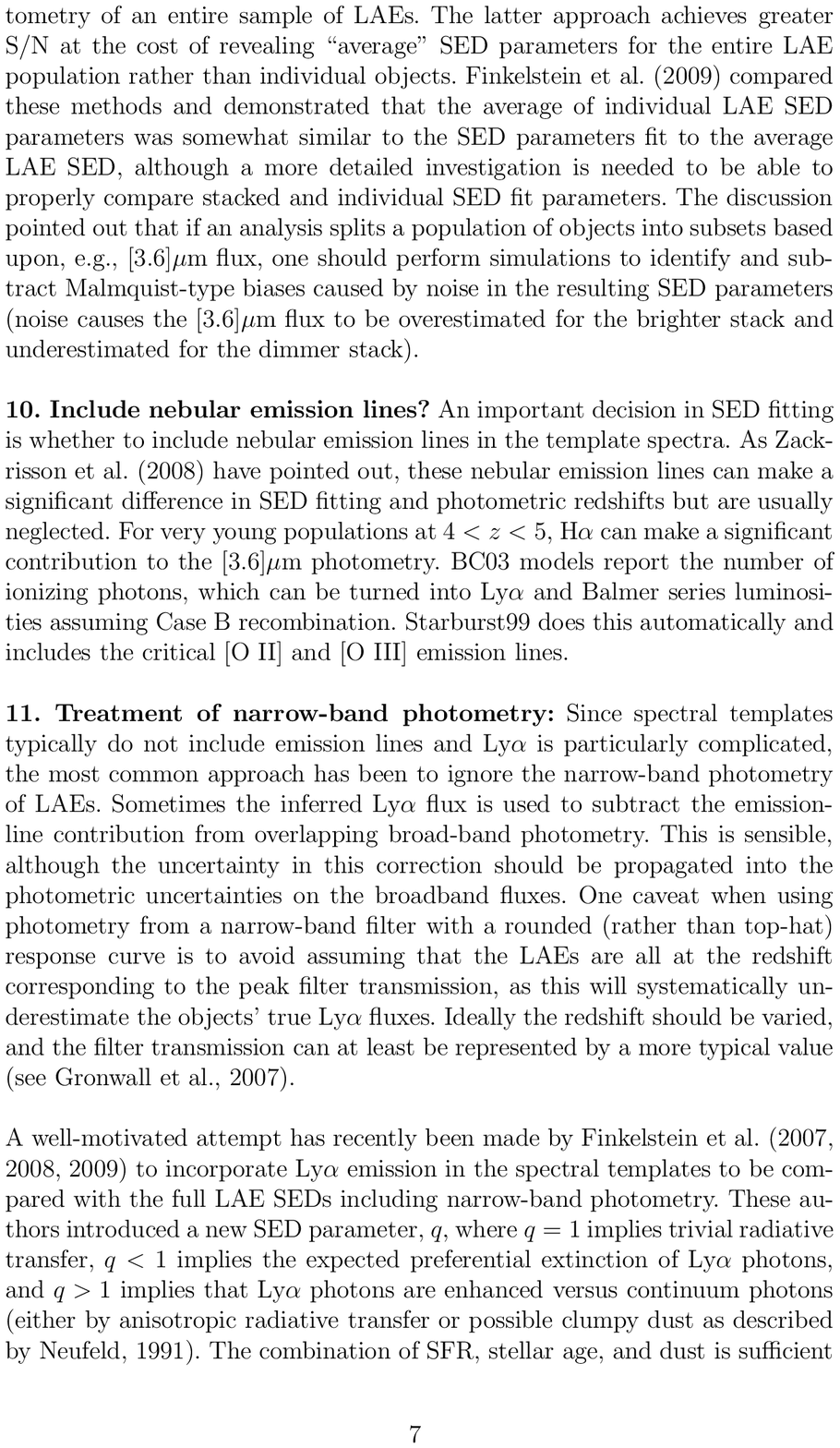,width=25cm,clip=}
\end{center}
\end{figure}
\newpage
\begin{figure}
\begin{center}
\epsfig{file=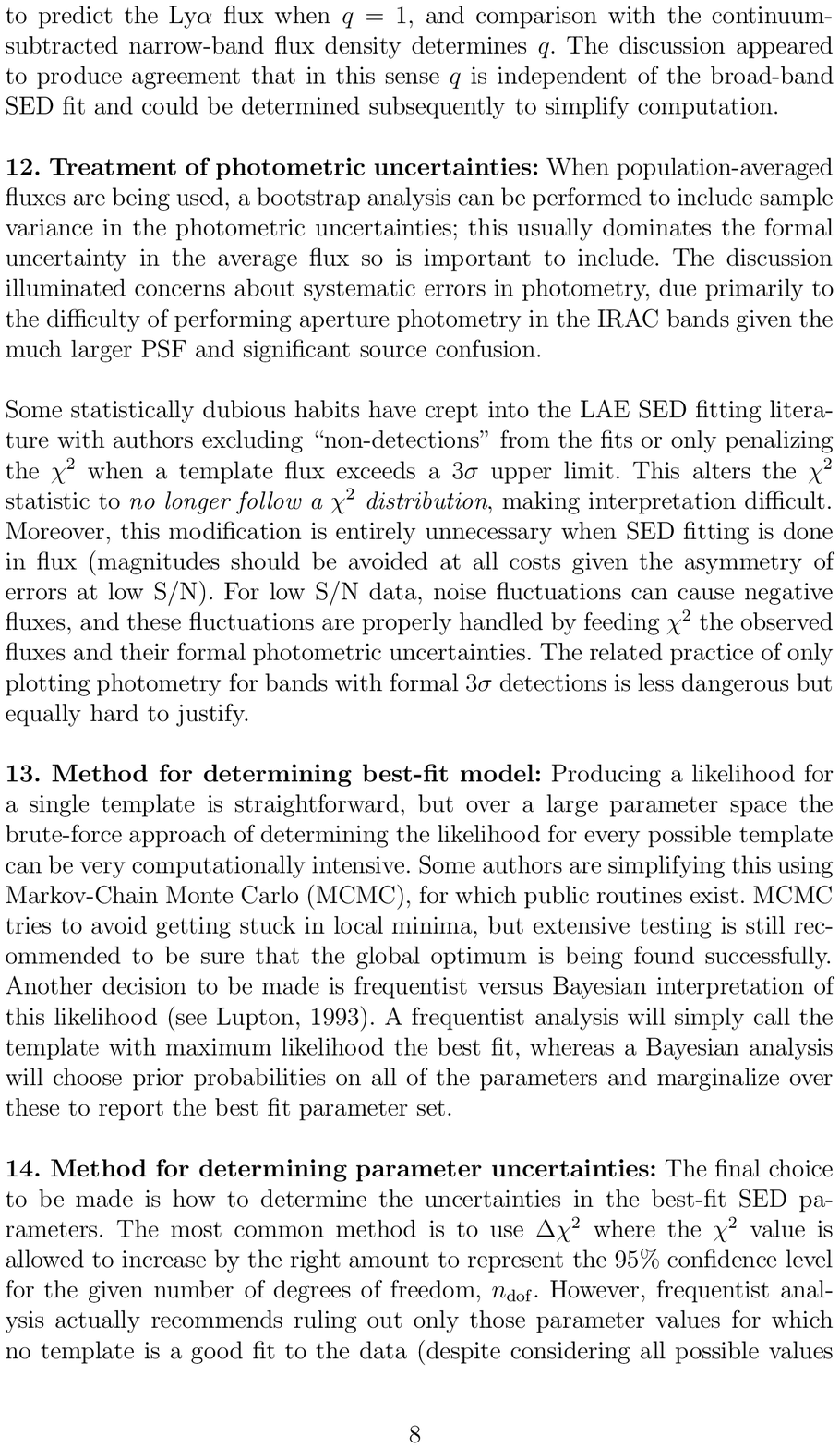,width=25cm,clip=}
\end{center}
\end{figure}
\newpage
\begin{figure}
\begin{center}
\epsfig{file=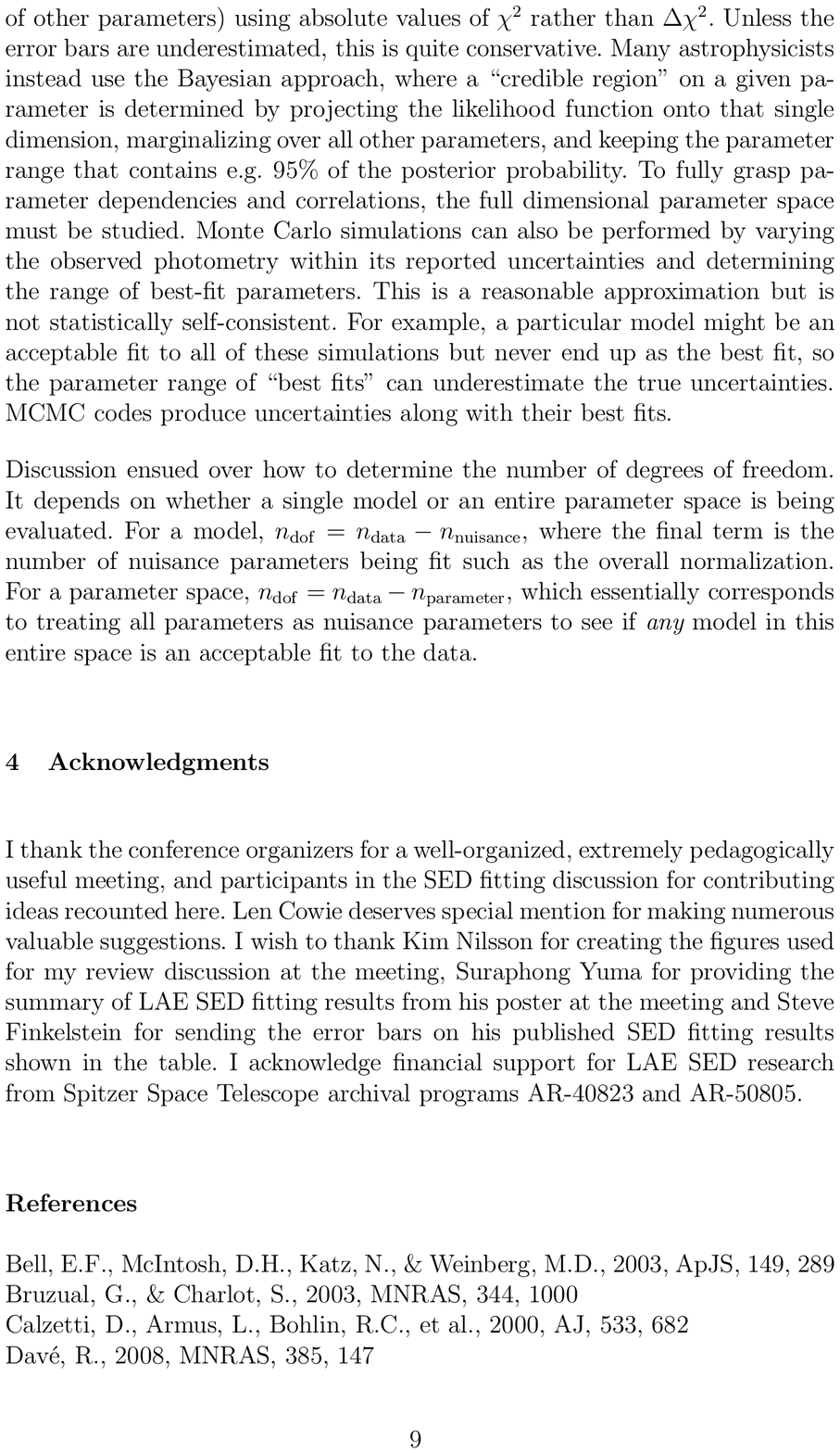,width=25cm,clip=}
\end{center}
\end{figure}
\newpage
\begin{figure}
\begin{center}
\epsfig{file=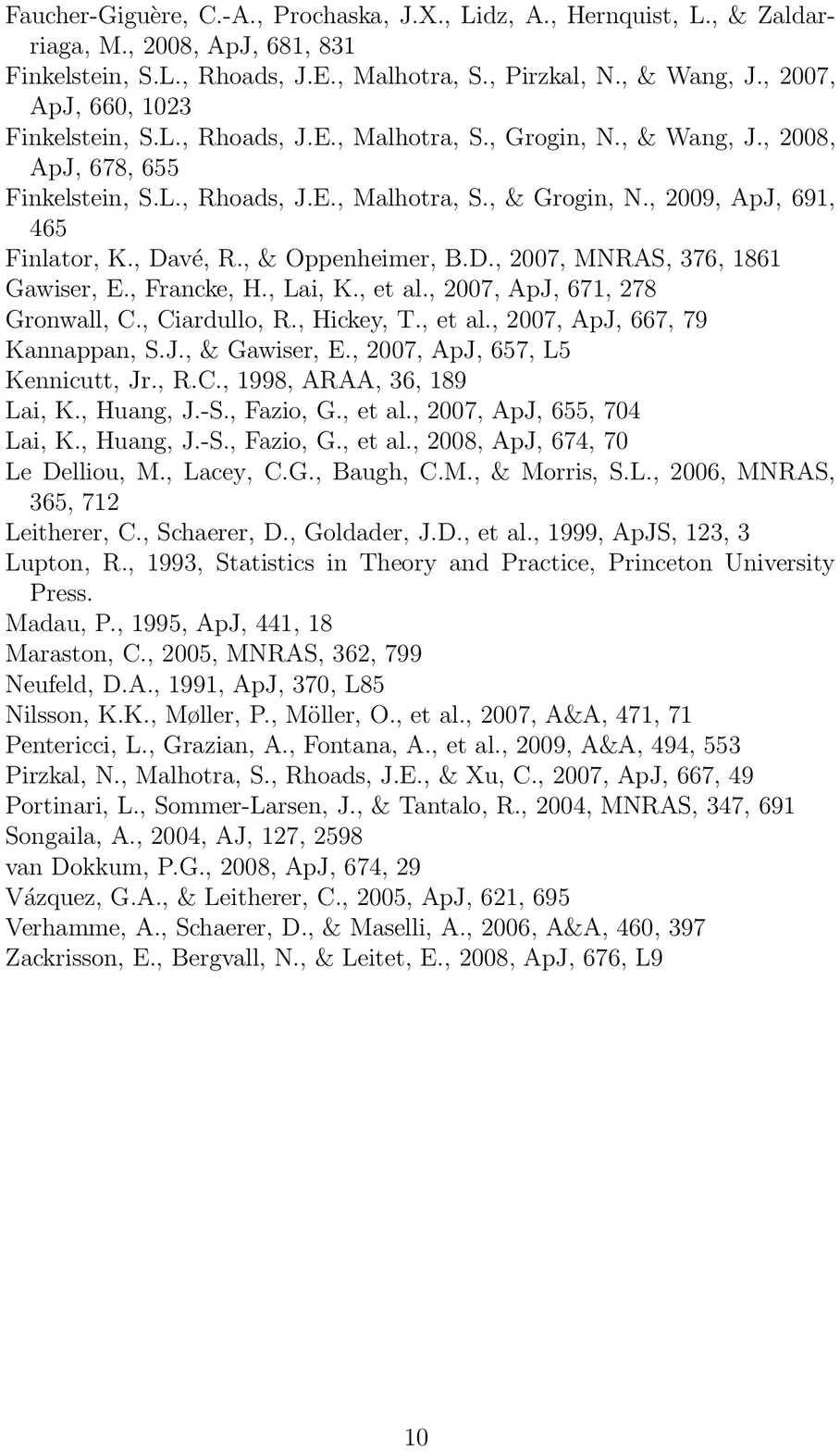,width=25cm,clip=}
\end{center}
\end{figure}
\newpage
\clearpage
\begin{figure}
\begin{center}
\epsfig{file=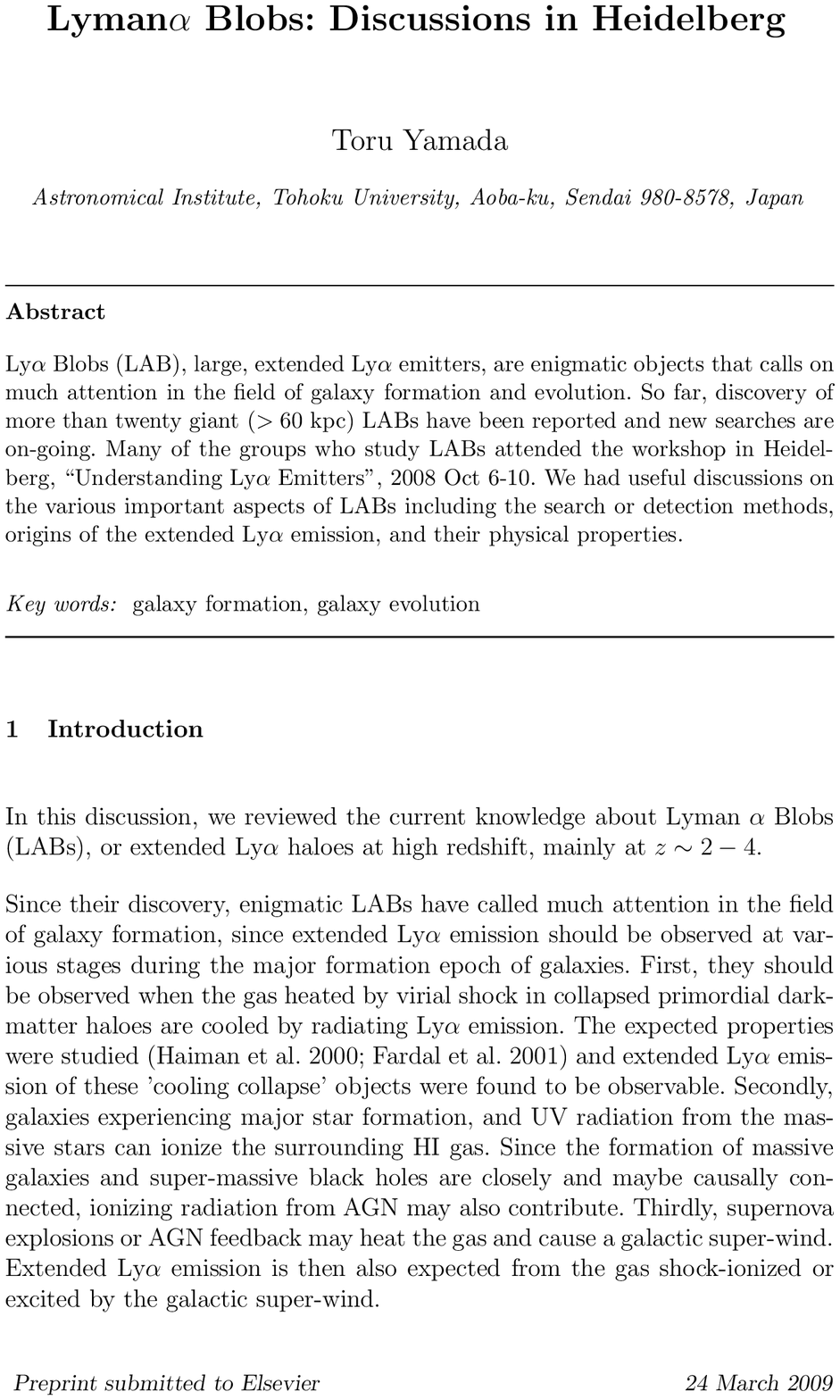,width=25cm,clip=}
\end{center}
\end{figure}
\newpage
\begin{figure}
\begin{center}
\epsfig{file=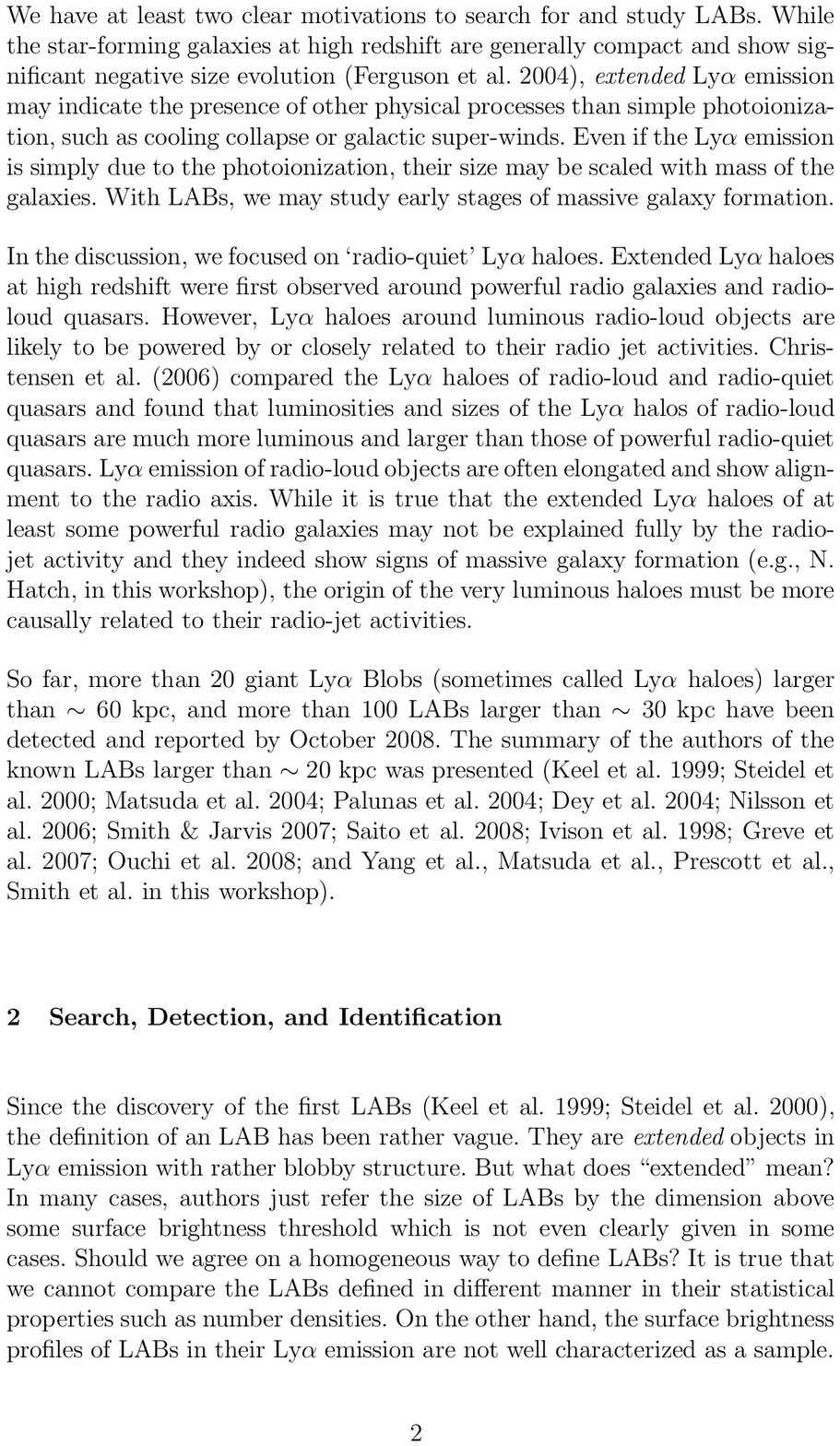,width=25cm,clip=}
\end{center}
\end{figure}
\newpage
\begin{figure}
\begin{center}
\epsfig{file=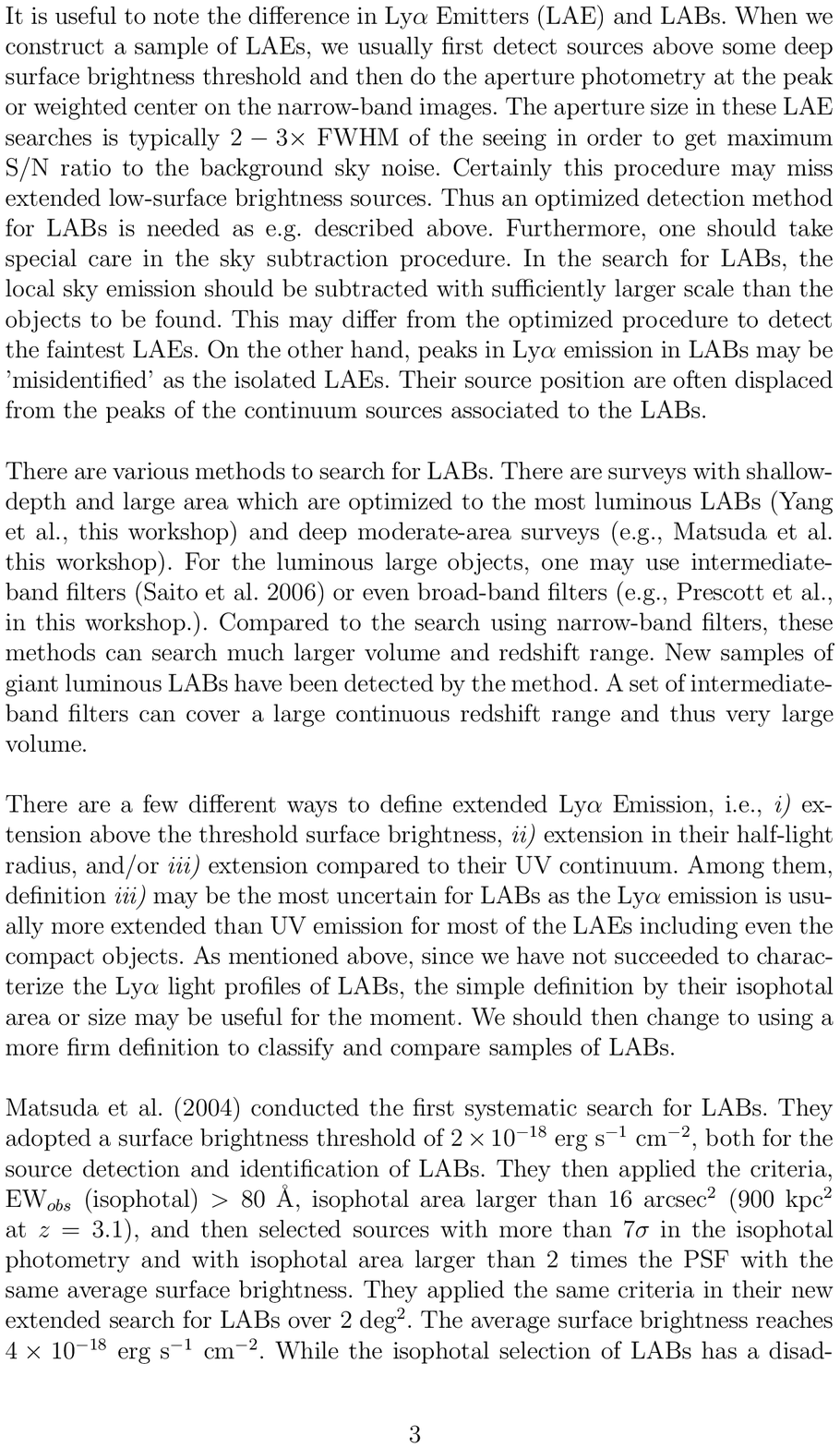,width=25cm,clip=}
\end{center}
\end{figure}
\newpage
\begin{figure}
\begin{center}
\epsfig{file=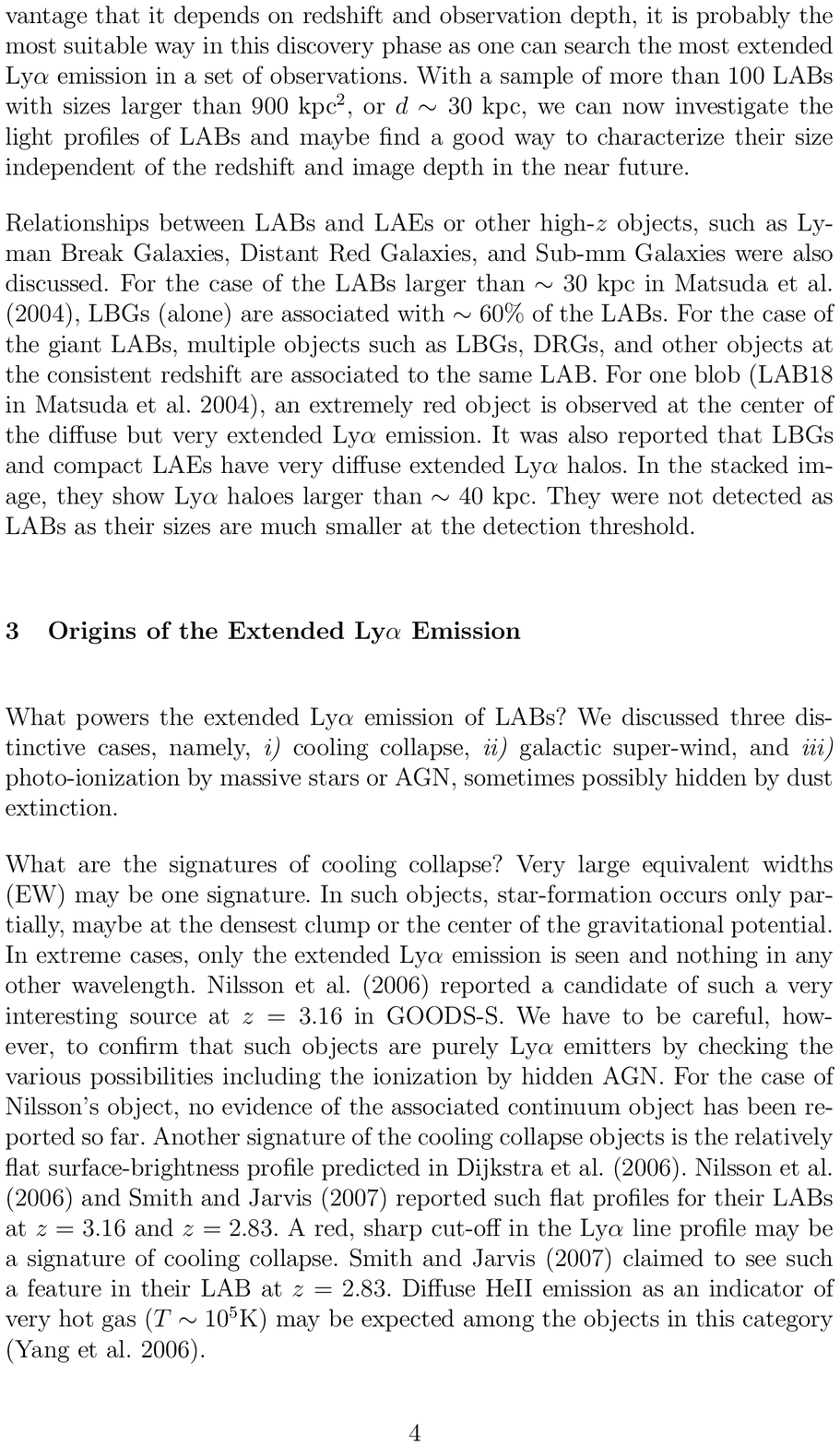,width=25cm,clip=}
\end{center}
\end{figure}
\newpage
\begin{figure}
\begin{center}
\epsfig{file=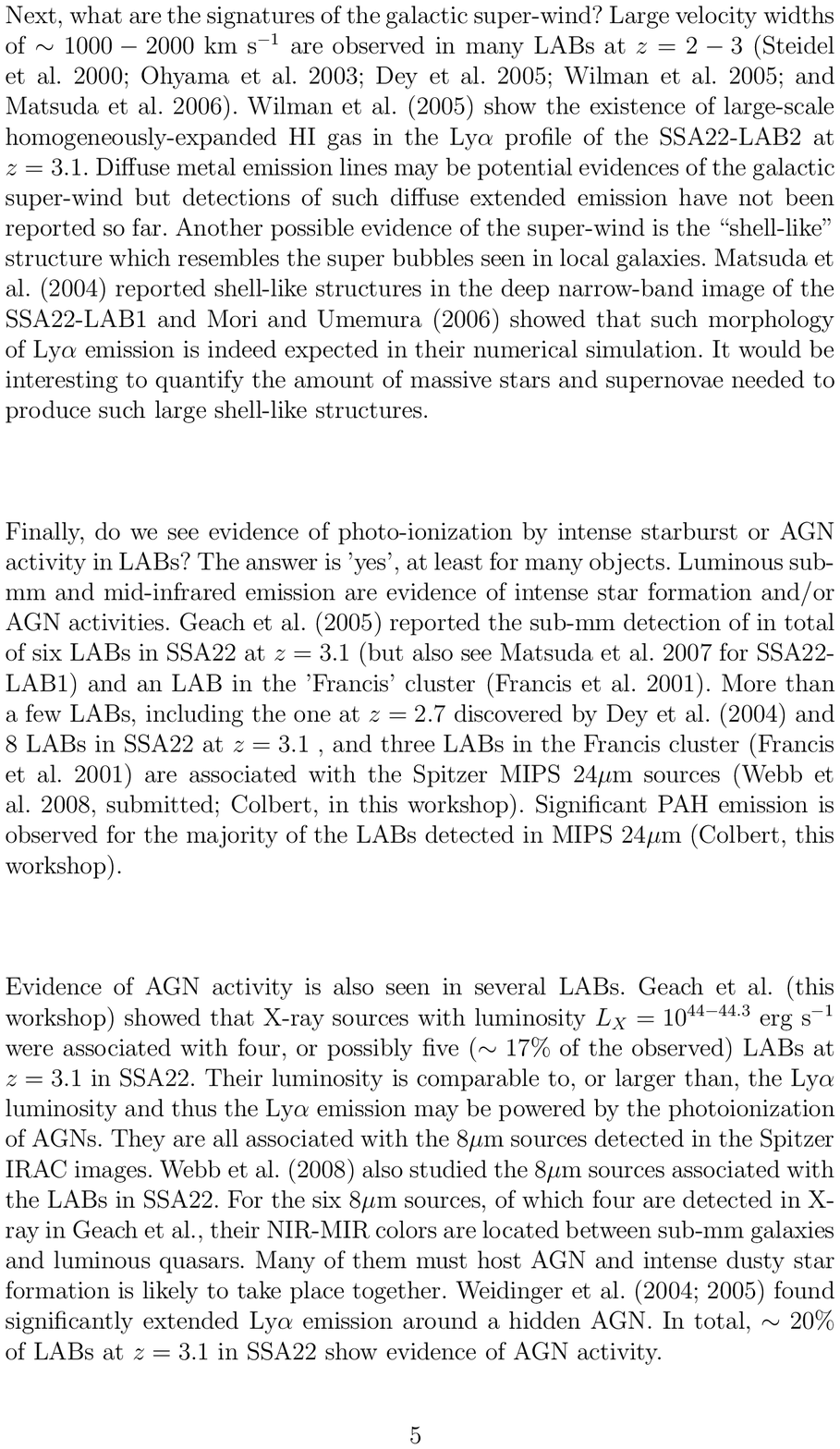,width=25cm,clip=}
\end{center}
\end{figure}
\newpage
\begin{figure}
\begin{center}
\epsfig{file=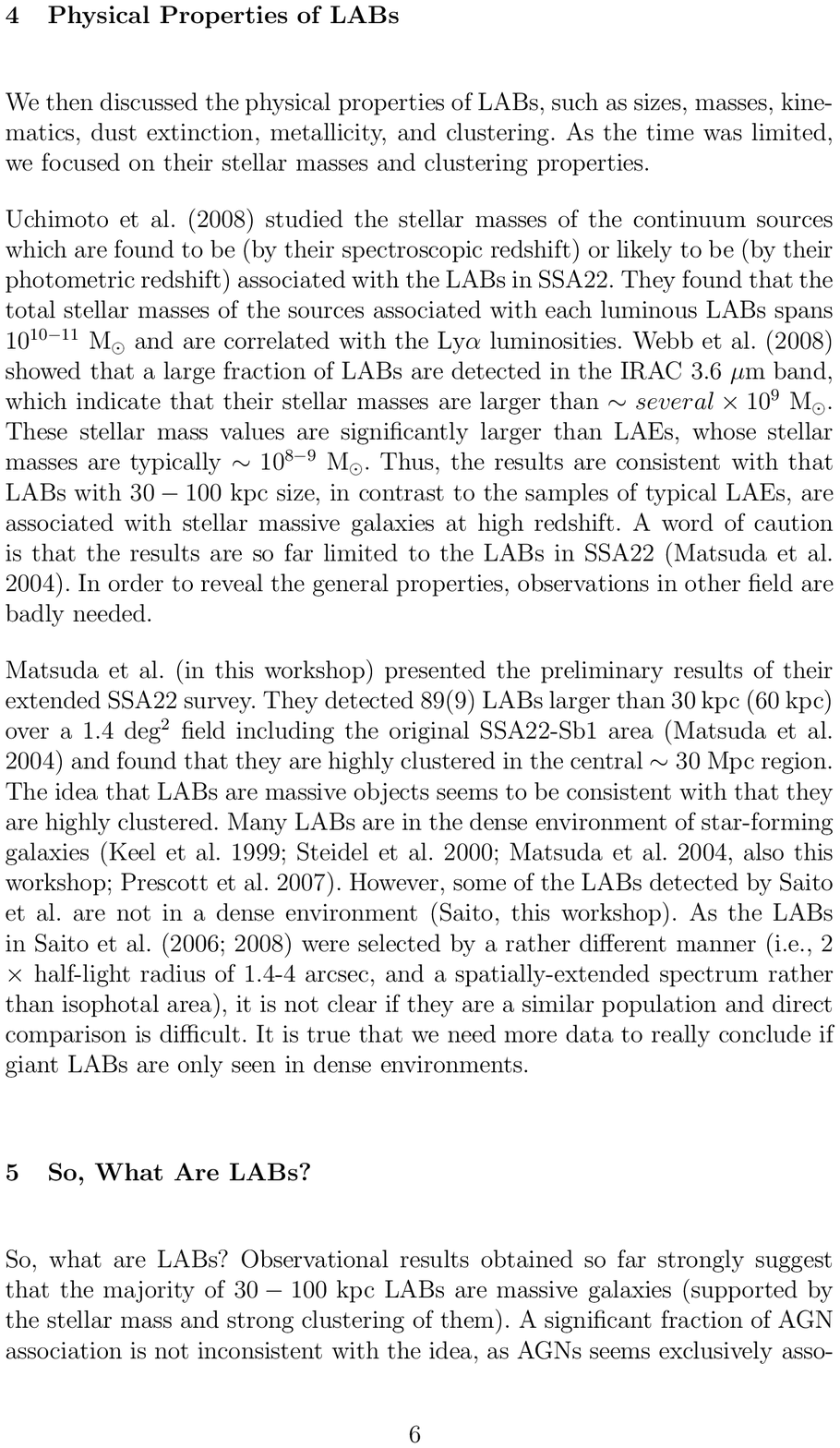,width=25cm,clip=}
\end{center}
\end{figure}
\newpage
\begin{figure}
\begin{center}
\epsfig{file=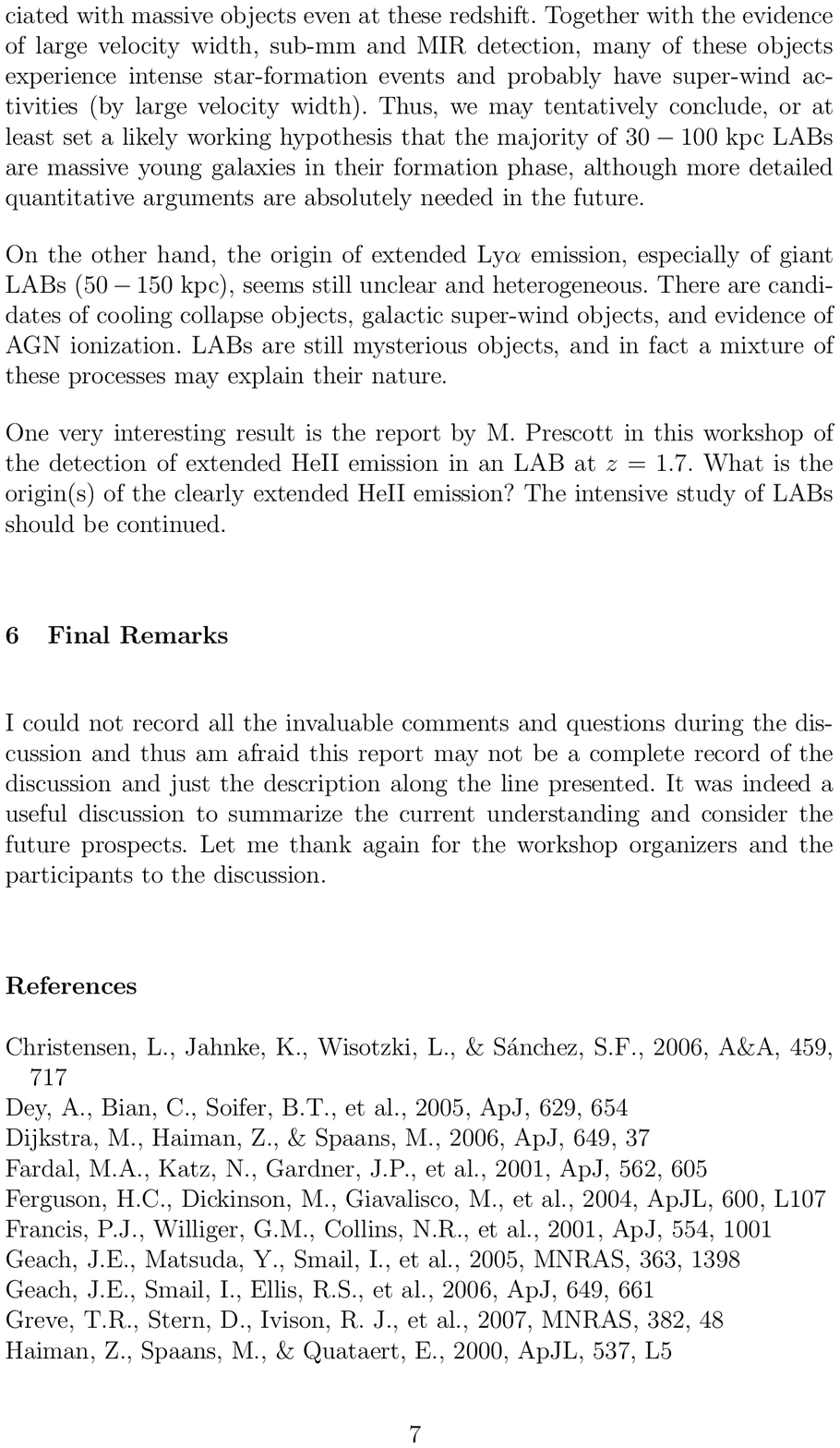,width=25cm,clip=}
\end{center}
\end{figure}
\newpage
\begin{figure}
\begin{center}
\epsfig{file=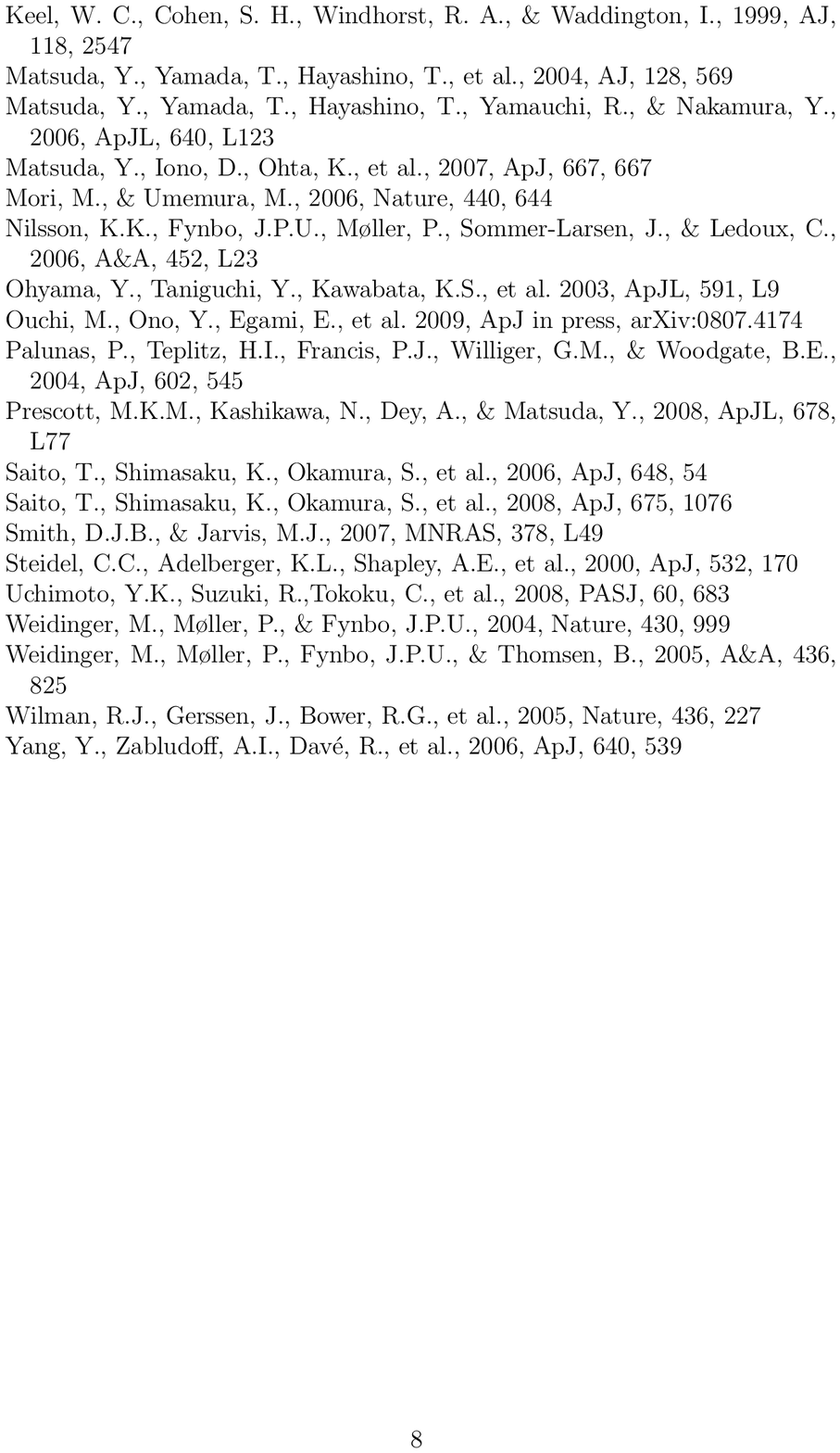,width=25cm,clip=}
\end{center}
\end{figure}

\end{document}